\documentclass[conference]{IEEEtran}
\IEEEoverridecommandlockouts

\usepackage{graphicx}
\graphicspath{{images/}}
\usepackage{amsthm}
\usepackage{subfig}
\usepackage{cite}
\usepackage{algorithm}
\usepackage{algpseudocode}
\usepackage{amsmath,amssymb,amsfonts}
\usepackage{graphicx}
\usepackage{textcomp}
\usepackage{xcolor}
\usepackage{booktabs}
\PassOptionsToPackage{linktocpage}{hyperref}
\usepackage[hyperindex,breaklinks]{hyperref}
\usepackage{cleveref}
\usepackage{breakurl}
\usepackage{threeparttable}
\usepackage{diagbox}
\newtheorem{lemma}{Lemma}[section]
\def\BibTeX{{\rm B\kern-.05em{\sc i\kern-.025em b}\kern-.08em
    T\kern-.1667em\lower.7ex\hbox{E}\kern-.125emX}}

\newcommand\vv[1]{\mathrm{#1}}

\DeclareMathOperator{\R}{\mathbb{R}}
\DeclareMathOperator{\timelockdelta}{\Delta^{\text{CLTV}}_e}

\usepackage{tikz}
\usetikzlibrary{arrows.meta}

\tikzset{mynode/.style={draw, very thick, circle, minimum size=1cm},
    myarrow/.style={very thick, -Triangle}}

\begin{document}

\title{An Exposition of Pathfinding Strategies Within Lightning Network Clients\\}

\author{\IEEEauthorblockN{Sindura Saraswathi}
\IEEEauthorblockA{\textit{Department of Computer Science} \\
\textit{University of North Carolina at Charlotte}\\
Charlotte, NC, USA \\
ssarasw2@charlotte.edu}
\and
\IEEEauthorblockN{Christian Kümmerle}
\IEEEauthorblockA{\textit{Department of Computer Science} \\
\textit{University of North Carolina at Charlotte}\\
Charlotte, NC, USA \\
kuemmerle@charlotte.edu}
}

\maketitle

\begin{abstract}
The Lightning Network is a peer-to-peer network designed to address Bitcoin's scalability challenges, facilitating rapid, cost-effective, and instantaneous transactions through bidirectional, blockchain-backed payment channels among network peers. Due to a source-based routing of payments, different pathfinding strategies are used in practice, trading off different objectives for each other such as payment reliability and routing fees. This paper explores differences within pathfinding strategies used by prominent Lightning Network node implementations, which include different underlying cost functions and different constraints, as well as different greedy algorithms of shortest path-type. Surprisingly, we observe that the pathfinding problems that most LN node implementations attempt to solve are NP-complete, and cannot be guaranteed to be optimally solved by the variants of Dijkstra's algorithm currently deployed in production. Through comparative analysis and simulations, we evaluate efficacy of different pathfinding strategies across metrics such as success rate, fees, path length, and timelock. Our experiments indicate that the strategies used by Eclair are advantageous in terms of payment reliability and result in paths with low fees. LND exhibits moderate success rates, while LDK results in paths with higher fee levels for smaller payment amounts; furthermore, CLN stands out for its minimal timelock paths. Additionally, we investigate the impact of Lightning node connectivity levels on routing efficiency. The findings of our analysis provide insights towards future improvements of pathfinding strategies and algorithms used within the Lightning Network.
\end{abstract}

\section{Introduction and Background} \label{sec:introduction}
Bitcoin \cite{nakamoto2008bitcoin}, the world’s first and arguably dominant decentralized digital currency, operates on a peer-to-peer network which allows transactions to be recorded securely on a distributed public ledger, also known as the blockchain, and to be verified individually, without the need for trusted third parties. However, Bitcoin encounters scalability challenges attributable to its existing block size limitations, which are challenging to overcome without compromising the decentralization or the security \cite{Buterin2017-Sharding,Nakai2024-Formulation} of the network. To address the scalability challenges inherent in Bitcoin, the Lightning Network (LN) \cite{poon2016bitcoin}, an off-chain payment channel network, was proposed in 2016.  

By leveraging the trust and security of the Bitcoin blockchain in its payment channels, the Lightning Network is designed to offer bitcoin-denominated fast, cost-effective, and scalable transactions. Within the LN, users establish bidirectional payment channels, which collectively form a peer-to-peer network. These channels operate as 2-of-2 multisignature contracts between two parties (nodes) that hold bitcoin and negotiate ownership of those coins. Through these channels, users, which serve as nodes within the LN, can conduct off-chain payments. Settlement of these off-chain payments on the Bitcoin blockchain occurs only when necessary, such as when a user opts to close a channel. This architecture renders the Lightning Network scalable and enables instant payments \cite{lightningOverviewBuilders}.

Two channel peers lock a certain amount of bitcoins into a 2-of-2 multisignature address, constituting the total capacity of the channel. Each peer possesses a portion of this total capacity, which they can spend. This spendable portion held by each channel peer is referred to as their channel balance or channel liquidity. As transactions occur within the channel, these balances may fluctuate. Depending on the direction of payments, one peer's balance will decrease while the other peer's balance increases \cite{antonopoulos2021mastering}.

The payment channel enables two nodes to conduct numerous transactions within the channel in both directions by generating revocable commitment transactions, thus eliminating the need for them to be broadcast on the blockchain \cite{di2018routing}. However, to securely route payments across multiple hops from the source to the destination, Hashed Timelock Contracts (HTLCs) are used \cite{poon2016bitcoin}.

Multi-hop payments are facilitated through source routing, wherein the source locates a path to the receiver for payment routing. Initially, the sender receives a cryptographic hash from the recipient. For the sender's payment to succeed, the recipient, upon receiving the HTLC, must return the pre-image along the HTLC-routed path within a specified timelock. Consequently, upon payment initiation, a series of HTLCs are established along the intended path. Each HTLC represents a commitment of funds from the preceding node to the next node, contingent upon the fulfillment of cryptographic proof. Once all contracts are established, the receiver shares the pre-image of the hash with its predecessor hop, which forwards it along the path until it reaches the sender, thereby resolving HTLCs along the path \cite{githubGitHubLightningbolts}. LN nodes can route payments only if they have sufficient balance on their side of the channel. A key practical challenge for successful payments within the LN is that the sender node selecting a payment path only possesses information about the capacity of channels, but not about the balance distribution within them \cite{EvaluatingPath,pickhardt2021security}, which might lead to the selection of payment path that cannot successfully route the payment amount.

In order to discover paths for routing payments, the sender node utilizes a pathfinding algorithm. Popular and widely used pathfinding algorithms have been developed by various LN implementations, including Lightning Network Daemon (LND) \cite{githubGitHubLightningnetworklnd}, Core Lightning (CLN) \cite{githubGitHubElementsProjectlightning}, Lightning Development Kit (LDK) \cite{githubGitHubLightningdevkitrustlightning,LDK_website}, and Eclair \cite{githubGitHubACINQeclair}. While mostly using a variant of a shortest-path algorithm, each of these node implementations differs in its pathfinding strategy in substantial details such as in the weight function that establishes underlying path costs. With their different choices, these LN client implementations adopt diverse approaches to the pathfinding process with different trade-offs regarding the fees incurred by payments, the typical length of payment paths, incurred delays and payment reliability. Only little research has been dedicated to selecting the appropriate client (or pathfinding strategy) based on different need profiles regarding these trade-offs inherent to payment delivery within LN. 

In this paper, we present a comprehensive analysis of the single-path pathfinding algorithms deployed by LND, CLN, LDK, and Eclair. Our objective is to compare the underlying channel weight functions, constraints, properties of deployed algorithms, and evaluate the performance and reveal the strengths and weaknesses of each client variant, providing insights for LN developers and researchers to guide future research and development efforts.

We discuss related works in \Cref{sec:relatedwork}, followed by an explanation of the pathfinding process within LN in \Cref{sec:Pathfinding}. In \Cref{sec:designelements}, we examine design elements of weight functions used in shortest path algorithms. We then present details on the pathfinding algorithms used by LN clients in \Cref{sec:beyond:shortest:path}, discussing also additionally imposed constraints, and provide a counterexample for the optimality of the modified Dijkstra variant used in LND for their imposed cost function. \Cref{sec:client:weights} delves into the discussion of implementation of pathfinding and weight functions used by different LN clients. We present our simulation model, experimental setup, results, and analysis in \Cref{sec:evaluation}. \Cref{limitations} discusses the limitations of our study. Finally, we conclude this paper in \Cref{conclusion}. 
\section{Related Work} \label{sec:relatedwork}
Numerous studies have examined off-chain payment channel networks like the Lightning Network, encompassing evaluations of network topology, enhancements to routing schemes, and considerations of security and reliability. 

The study by \cite{pickhardt2021security} concentrates on the reliability and privacy of LN payments by introducing a mathematical framework to model uncertainty in channel balances using probability theory. Furthermore, \cite{kappos2021empirical} explores the privacy properties of the Lightning Network through simulations. \cite{tochner2020route} investigates the structure of the Lightning Network and the routing algorithms used by LND, CLN and Eclair. Then it explores the effects of routing choices by analyzing the feasibility of Denial-of-Service attacks and route hijacking across these three implementations. In our work, we specifically analyze the effects of design elements in the pathfinding algorithm and evaluate their empirical performance from an operational perspective.

\cite{kotzer2023braess} examines the potential presence of the Braess paradox in off-chain networks and proposes solutions to mitigate its impact through analysis employing three routing algorithms, including LND and CLN. \cite{camilo2024profitpilot} analyzes the Lightning Network's topology and introduces a strategy called ProfitPilot for connecting new nodes in payments by encouraging cycle creation in payment channel networks. Studies such as \cite{camilo2022topological,kappos2021empirical,guo2019measurement,Guasoni-2024Lightning} present topological analyses of the Lightning Network. In our study, we utilize insights into the Lightning Network topology, including the node connectivity levels, to analyze the performance of pathfinding strategies for different network participants.

While our research focuses on existing implementations of single-path pathfinding strategies within the current LN protocol, several studies have proposed modifications to the routing protocol that go beyond the current capabilities of LN. Flare \cite{prihodko2016flare} proposes a routing methodology in the Lightning Network that relies on routing tables and beacon nodes to determine paths. Spider \cite{sivaraman2020high} presents a routing solution that segments transactions into packets and utilizes a multi-path transport protocol to ensure balanced channel usage. Boomerang \cite{bagaria2020boomerang} introduces redundant payments on existing multi-path routing schemes to reduce latency. 

Beyond the scope of our exposition are payment delivery strategies based on (atomic) multi-part payments \cite[Chapter 12]{antonopoulos2021mastering}, which have been introduced as an extension to the original LN protocol \cite{githubBaseRustyrussell}. \cite{pickhardt2021optimally} argues that multipart payment delivery can benefit from a minimum cost flow formulation, which generalizes the concepts of payment paths, instead of following a shortest-path framework. \cite{rahimpour2021spear} proposes a variation of Boomerang which requires less computing power and latency for faster multi-path payments. While multi-part payments are very promising, they do currently not (yet) represent the mainstream practice in all LN implementations, where single-path payments still dominate. Thus, our study focuses on single-path payments, but the conducted analysis could be extended to multi-part payments in future work.

A paper with a similar scope as our exposition is \cite{kumble2021comparative}, which undertakes a comparative examination of routing clients, analyzing weight functions of LND, CLN, and Eclair. However, significant developments and enhancements have occurred in these routing clients since that study, and \cite{kumble2021comparative} does not consider constraints or modification of Dijkstra's algorithm which are necessary to account for certain choices of cost functions, e.g., the ones used by LND. As an additional routing client, our research also includes LDK alongside these clients. The distinct experimental methodology to evaluate the performance of different pathfinding strategies is further different than that of \cite{kumble2021comparative}, as we also consider different connectivity levels of nodes. Finally, the  systematic analysis of weight function design elements as presented in \Cref{sec:designelements} is novel to the best of our knowledge.
\section{Pathfinding and Route Selection} \label{sec:Pathfinding}
For the purpose of pathfinding, the Lightning Network can be modeled as a directed graph $G=(V,E)$, in which each network participant is represented as a node or vertex $v \in V$, and a payment channel between two nodes $u,v \in V$ is represented as the pair of edges $e=(u,v) \in E$ and $e=(v,u) \in E$. Each edge in the graph is associated with specific channel attributes such as fees, capacity, timelock delta, age, and information amount the maximal and minimal amount of currency units that can be settled in an HTLC. These channel attributes are also utilized as design elements for pathfinding purposes. It is important to note that the LN graph evolves dynamically over time--the number of nodes and channels changes through channel opening and closing transactions, and beyond that, node operators adapt channel parameters such as fee parameters and communicate those changes via a gossip protocol to other network participants.

In LN transactions, the node that initiates the payment is often referred to as the source node, while the node that receives the payment is called the destination or receiver node. There are several intermediary nodes that facilitate the forwarding of payments between the source and receiver nodes. These intermediary nodes are commonly referred to as routing nodes, and the channels included in this payment forwarding are referred to as hops.

When the payment is initiated, the source node needs to select a path to the receiver node using pathfinding strategies on the LN graph. Existing LN clients employ instantiations of shortest path algorithms, such as Dijkstra's algorithm, modified Dijkstra-style algorithms or Yen's $K$-shortest path algorithm, to find the path between specified source and receiver nodes \cite{githubLndroutingpathfindgoMaster}, \cite{githubLightningcommondijkstracMaster}, \cite{githubRustlightninglightningsrcroutingrouterrsMain} and \cite{githubEclaireclaircoresrcmainscalafracinqeclairrouterGraphscalaMaster}. We refer to \Cref{sec:beyond:shortest:path} for a discussion of these variants in the context of their use in LN implementations.

\section{Weight Function Design Elements} \label{sec:designelements}
Given a sender node $s \in V$ and a recipient node $r \in V$, shortest path algorithms aim to find a path $p = (e_1,\ldots,e_{\operatorname{length}(p)})$ of edges $e_i \in E$, where $s$ is the initial node of $e_1$ and $r$ is the terminal node of $e_{\operatorname{length}(p)}$, which minimizes a total path cost $c(p)$ given as the sum of \emph{edge weights} given by a function $\operatorname{weight}: E \to \R_{\geq 0}$ that maps to non-negative real numbers $\R_{\geq 0}$, such that
\begin{equation} \label{eq:additive:pathcost}
    c(p) := \sum_{i=1}^{\operatorname{length}(p)} \operatorname{weight}(e_i) = \sum_{e \in p} \operatorname{weight}(e),
\end{equation}
cf. \cite[Chapter 22]{Cormen-2022Introduction}. Recalling that an edge in the LN graph corresponds to a payment channel, this means that current LN client implementations, which use the shortest path algorithms to find their payment paths (see \Cref{sec:beyond:shortest:path}), implicitly assume additive path costs in the sense of \eqref{eq:additive:pathcost}.

Depending on different design choices that are made, the edge weight function $\operatorname{weight}(\cdot)$ is composed of different terms, which try to capture different desired properties of the resulting payment path returned by the pathfinding algorithm. Specifically, in the following, we model the edge weight function to have the form
\begin{equation} \label{eq:edgeweight:def}
\operatorname{weight}(e) = \sum_{i=1}^k c^i a_e^i, 
\end{equation}
where $k$ is a number of distinct elementary weight terms, $a_e^i \geq 0$ is an edge-specific factor associated to the $i$-th weight term, and $c^i \geq 0$ is an associated factor or constant that does not depend on the edge $e$ in consideration. 

However, the choice of these factors $c^i$ and $a_e^i$ varies across the clients. In this section, we examine various design components and concepts contributing to weight functions of LN clients, with a specific focus on discussing the principles and model assumptions underlying these components. The usage of each of these elements can lead to payment paths exhibiting a variety of different properties.
\subsection{Fees} \label{sec:fees}
When routing payments from a source $s$ to a recipient node $r$, the payment path $p$ typically involves intermediary nodes and multiple hops. These intermediary nodes charge fees for each successfully settled payment they participate in.
These fees can be subdivided into two distinct parts: the base fee ($\vv{b}_e$), which remains constant regardless of the transaction size, and the proportional fee or fee rate ($\vv{f}_e$), which varies based on the amount being forwarded through the node. Consequently, the total fee charged by an intermediary node for forwarding a specific payment amount ($\vv{amt}_e$) is determined by the combination of these two components, such that
 \begin{equation}\label{one}
    \vv{fee}_e = \vv{b}_e + \vv{amt}_e \cdot \vv{f}_e.
\end{equation}
Since it is reasonable for the source node to try to find a path with fees as low as possible, one of the weight terms $a_e^i$ in \eqref{eq:edgeweight:def}
can be chosen as $a_e^i = \vv{fee}_e$ defined above. This choice is a primary design consideration in all LN implementations when formulating their weight functions. Since fees are directional, i.e., they can vary depending on the direction of payment flow between the same two nodes $u,v \in V$, a modeling of the LN as a directed graph instead of an undirected one is more appropriate. We also note that a purely edge-dependent fee as modeled in \eqref{one} is not fully realistic, as the payment amount $\vv{amt}$ needs to cover fees charged by routing nodes along the selected path, making $\vv{amt}$ path-dependent. Since the fees are typically several orders of magnitudes smaller than the amount to be received by the receiver node, we neglect this effect, in accordance with the practice in current LN node implementations.
\subsection{Timelock}
Once a specific number of blocks is reached---referred to as a \emph{timelock}---each incoming HTLC in a channel expires. This timelock serves as a time buffer within which the node aims to complete all the HTLC exchanges for a given transaction. For the length of the timelock, funds are locked in the payment channels.
The total timelock of an HTLC, in direction from sender to recipient, depends on the timelock of the subsequent HTLC along a payment path plus the timelock difference value $\Delta^{\text{CLTV}}_e$, called \emph{timelock delta} or \emph{CLTV expiry delta}\footnote{CLTV is an acronym of the Bitcoin script operator OP\_CHECKLOCKTIMEVERIFY \cite[Glossary]{antonopoulos2021mastering}.} that is set by the node $v$ at the outgoing side of the channel $e = (u,v)$ \cite[Chapter 8]{antonopoulos2021mastering}. In order to ensure that funds are locked for a shorter duration, it is reasonable for routing clients to prioritize the inclusion of channels with lower timelock deltas into the payment path compared to channels with higher timelock deltas.

For this reason, to penalize channel with large timelock deltas, a weight term $a_e^i$ in \eqref{eq:edgeweight:def} can be chosen to be proportional to the timelock delta $\Delta^{\text{CLTV}}_e$. In LND, CLN, and Eclair, an edge-dependent proportionality factor is chosen such that $a_e^i = \vv{amt}_e \cdot \Delta^{\text{CLTV}}_e$. The corresponding constant, potentially tunable factor $c_i$ is called $\vv{riskfactor}$ in LND and CLN and $\vv{lockedFundsRisk}$ in Eclair. In LDK, an upper bound on the timelock delta is used as a constraint for a channel to be considered during pathfinding, but it does not appear as part of a weight term. 
\subsection{Success Probability} \label{ele:successprob}
Due to the fact that payment paths in which the desired payment amount can be settled reliably are preferred, it is common to take an estimate of a success probability, which estimates the likelihood of a successful payment through a payment path, into account when searching for suitable payment paths $p$. In this subsection, we first discuss two different approaches for incorporating payment path success probabilities into a model for a total path cost, and then shed light on different approaches for modeling individual channel success probabilities. 
\subsubsection{Path Success Probability} \label{ele:pathprob}
Given a path $p$ between sender node $s \in V$ and recipient node $r \in V$, assume that we have access to an estimate $P_{p} \in [0,1]$ for its success probability. 

In existing literature on LN pathfinding, there are two different approaches for incorporating a path success probability estimate $P_p$ into a total path cost $c(p)$ of a payment path $p$---one with an additive term inversely proportional to $P_p$ and the second proportional to the negative logarithm of $P_p$.

The first, which is used in LND \cite{inverseProbDerivation,lightningClearingPaths} implies a success probability penalization that is proportional to the inverse $P_p^{-1}$ of the path success probability $P_p$. It is based on the following lemma, the proof of which is provided in \cite{lightningClearingPaths}.

\begin{lemma}
Assume that we compare two paths $a$ and $b$ between sender $s \in V$ and recipient $r \in V$ which incur a fee cost (e.g., incurred routing fee) of $F_a$ and $F_b$, respectively, if successfully used to route the payment. Assume that an additional fixed virtual cost $c_{\text{attempt}}$ is incurred for a payment attempt along any path $p$.\footnote{See also \Cref{ele:attemptcost}.} Then, the expected total cost after two payment attempts is minimized by trying first the path $p$ which is a minimizer of the objective
\begin{equation} \label{eq:cost:inversePp}
c(p) = F_p + \frac{c_{\text{attempt}}}{P_p} + c_0,
\end{equation}
where $c_0 \in \R$ is an arbitrary, path-independent constant.
\end{lemma}
\begin{proof}
    If path $a$ is tried and succeeds, a cost of $F_a + c_{\text{attempt}}$ is incurred. On the other hand, if path $a$ is tried and does not succeed, there are two possible outcomes: Either the subsequently path $b$ succeeds, in which case a total cost of $F_b + 2 c_{\text{attempt}}$ is incurred, or subsequent path $b$ fails, in which a total cost of $ 2 c_{\text{attempt}}$ is incurred.

    The cost $c_{ab}$ of attempting to use first path $a$ and then $b$ is thus in expectation
    \[
    \begin{split}
    \mathbb{E}[c_{ab}] &= P_{a} (F_a + c_{\text{attempt}}) + (1-P_{a}) P_b (F_b + 2 c_{\text{attempt}}) \\
    &+ (1-P_{a})(1-P_{b}) 2 c_{\text{attempt}},
    \end{split}
    \]
    whereas the cost $c_{ba}$ of attempting to use first path $b$ and then $a$ is in expectation
    \[
    \begin{split}
    \mathbb{E}[c_{ba}] &= P_{b} (F_b + c_{\text{attempt}}) + (1-P_{b}) P_a (F_a + 2 c_{\text{attempt}}) \\
    &+ (1-P_{b})(1-P_{a}) 2 c_{\text{attempt}}.
    \end{split}
    \]
    In both cases, the outcome regarding payment success is the same--if any of the two paths $a$ and $b$ is able to route the payment successfully, the payment will be successful, otherwise, the payment has failed. Thus, it is sufficient to compare $\mathbb{E}[c_{ab}]$ with $\mathbb{E}[c_{ba}]$ to obtain a preference for $a$ over $b$ or vice versa. In particular, we see that $\mathbb{E}[c_{ab}] < \mathbb{E}[c_{ba}]$ is equivalent to 
    \[
    \begin{split}
    &P_{a} c_{\text{attempt}}  - 2 P_{a}P_{b} c_{\text{attempt}} -P_{a}P_{b}F_{b} + 2 P_b c_{\text{attempt}}   \\
    &<P_{b} c_{\text{attempt}} - 2 P_{a}P_{b} c_{\text{attempt}} -P_{a}P_{b}F_{a} + 2 P_a c_{\text{attempt}},
    \end{split}
    \]
    which holds if and only if
    \[
    \begin{split}
    &P_{a}P_{b}F_{a} + P_b c_{\text{attempt}}   \\
        &< P_{a}P_{b}F_{b} + P_a c_{\text{attempt}}.
    \end{split}
    \]    
If neither success probability $P_a$ nor $P_b$ is $0$, via multiplication by $\frac{1}{P_{a}P_{b}}$, this is equivalent to
\[
F_{a} + \frac{c_{\text{attempt}}}{P_a} + c_0 < F_{b} + \frac{c_{\text{attempt}}}{P_b} + c_0,
\]
for any $c_0\in \R$, which shows that choosing $p \in \{a,b\}$ minimizing \eqref{eq:cost:inversePp} leads to the desired outcome. If either $P_a = 0$ or $P_b = 0$, minimizing $\eqref{eq:cost:inversePp}$ also leads to the desired outcome as \eqref{eq:cost:inversePp} corresponds to an infinite expected total cost for $P_p = 0$, but a finite expected total cost for any path with non-zero $P_p$.
\end{proof}
We note that in \cite{inverseProbDerivation,lightningClearingPaths}, the above lemma is derived assuming the existence of a virtual cost $c_{\text{attempt}}$ only in case of a payment failure, but both assumptions lead to the same objective \eqref{eq:cost:inversePp}. In \eqref{eq:cost:inversePp}, the fee cost term $F_p$ fits well into the additive framework of \eqref{eq:additive:pathcost} due the fact that routing fees accumulate additively along the path such that $F_p = \sum_{e \in p} \operatorname{weight_a}(e)$, where $\operatorname{weight_a}: E \to \R_{\geq 0}$ is a non-negative, additive weight function. 

However, for second summand of \eqref{eq:cost:inversePp}, the situation is different as an additive decomposition of $\frac{c_{\text{attempt}}}{P_p}$ across edges can be mathematically not justified. On the other hand, a reasonable assumption is that a payment across a path $p$ is successful if and only if the routing across \emph{all} channels $e \in p$ is successful, who themselves are assigned channel success probabilities $P_e$ \cite{pickhardt2021security}. If we assume that the routing success across one channel $e \in p$ is independent from the routing success across any other channels along each path, we have that
\begin{equation} \label{lnd:probpath}
\begin{split}
P_{p} &= P\left(\{\text{payment across path } p \text{ is successful}\}\right) \\
&= P\left(\cap_{e \in p} \{\text{payment across channel } e \text{ is successful}\}\right) \\
&= \prod_{e \in p} P_e,
\end{split}
\end{equation}
using the independence assumption in the last equality. In view of this equality, it is possible to decompose the cost of \eqref{eq:cost:inversePp} across edges if we introduce the multiplicative weight $\operatorname{weight_m}: E \to \R_{\geq 1}, e \mapsto 1/P_e$ such that, in the case of $c_0 =0$,
\begin{equation} \label{eq:mixed:additivemult:cost}
c(p) = F_p + \frac{c_{\text{attempt}}}{P_p} =  \underbrace{\sum_{e \in p} \operatorname{weight_a}(e)}_{=:c_a(p)} + \underbrace{c_{\text{attempt}} \prod_{e \in p} \operatorname{weight_m}(e)}_{=:c_m(p)}.
\end{equation}

Unfortunately, it is not possible to use a standard shortest path algorithm such as Dijkstra's algorithm to find the path with the lowest additive-plus-multiplicative cost \eqref{eq:mixed:additivemult:cost} as these algorithms all assume additive cost as in \eqref{eq:additive:pathcost} \cite[Chapter 22]{Cormen-2022Introduction}.
The LND LN node implementation mitigates this issue by modifying the priority queue used in Dijkstra's algorithm to be adapted to \eqref{eq:mixed:additivemult:cost}. However, as we discuss in \Cref{sec:beyond:shortest:path}, it can be shown that such a modified Dijkstra is not guaranteed to find the optimal path with respect to \eqref{eq:mixed:additivemult:cost}. 

A second approach to incorporate the path success probability $P_p$ into a path cost function is to add a factor proportional to $\log\left(\frac{1}{P_p} \right)$ instead of $\frac{1}{P_p}$. Under the independence assumption of the payment across path edges $e$ mentioned above, we see that
\begin{equation}\label{eq:log_prob}
\begin{aligned}
    \log \left(\frac{1}{P_{p}}\right) &= -\log\left(\prod_{e \in p}P_{e}\right) &= -\sum_{e \in p} \log (P_{e}),
\end{aligned}
\end{equation}
which leads to a fully additive cost of the type \eqref{eq:additive:pathcost} motivates the choice of a weight term $a_e^i = - \log(P_{e})$ in \eqref{eq:edgeweight:def}. Since the edge success probabilities $P_e$ are between $0$ and $1$, these weight terms are always non-negative. If used without any other weight terms (e.g., fee-related ones), the resulting problem type corresponds to finding a \emph{maximally reliable path} and is known to be equivalent to a shortest path problem \cite[Exercise 4.39]{Ahuja-1993Network}. CLN, LDK and Eclair use this logarithmic success probability term in their weight function to prioritize paths with higher success probability within their pathfinding algorithms.

Comparing the two approaches, we note that while an inverse penalization $P_p$ leads to difficulties in the pathfinding algorithms as mentioned above, an argument can be made that this modeling is more suitable to avoid unreliable paths as $1/P_p$ grows faster for $P_p \to 0$ than $-\log(P_p)$, i.e., the penalization of unreliable vs. reliable paths is stronger in the modeling used by LND.

\subsubsection{Channel-Wise Success Probability} \label{ele:edgeprob}
We discussed in the last subsection how, through considering path success probabilities and a suitable independence assumption, weight terms based on $P_e$, the probability that the channel $e$ is able to successfully route a payment, are suitable design elements within total path costs \eqref{eq:additive:pathcost}. There are different approaches to estimate this channel-wise success probability $P_e$, underlying different assumptions. 

As a simple, first approach, every untried channel can be assumed to have a constant default success probability \cite{lightningClearingPaths}. For channels used in routing payments historically, the success probability estimation can be based on the outcomes of historical payment attempts \cite{githubGitHubLightningnetworklnd}. If historical payment data is collected across many payment attempts, in the event of a failed payment attempt, the channel success probability estimate $P_e$ can be decreased to $0$. The channel success probability estimate $P_e$ gradually returns to its default value over time.

In LND, there are two ways to calculate the channel's success probability: $Apriori$ and $Bimodal$. This approach is followed in LND's $Apriori$ mode.

Another popular model to estimate $P_e$ is based on modeling the relationship between the intended payment amount and the channel capacity \cite{pickhardt2021security,githubGitHubACINQeclair}. Capacity is the sum of balances across the nodes in the channel. While channel's capacity is publicly known, the individual balances remain private. A large channel capacity relative to the payment amount intuitively leads to greater probability of success, which is why it can be used within a success probability estimate. Specifically, assuming a uniform distribution for channel liquidity, the failure probability can be expressed as the ratio of payment amount $\vv{amt}_e$ to the channel capacity $\vv{cap}_e$, such that the success probability $P_e$ can be written as 
\begin{equation} \label{eq:uniform:modeling}
P_e = 1- \frac{\vv{amt}_e}{\vv{cap}_e} = \frac{\vv{cap}_e - \vv{amt}_e}{\vv{cap}_e}.
\end{equation}
In this modeling, the success probability $P_e$ decreases as the payment amount approaches the channel's capacity. This model is used in the Eclair LN node implementation \cite{githubGitHubACINQeclair}.

Based on the same assumption, a finer model for $P_e$ can be obtained if upper and lower bounds $\vv{UB}_e$ and $\vv{LB}_e$ for the available balance (or liquidity) in a channel $e$ are available, as mentioned in \cite{githubGitHubLightningdevkitrustlightning}. This approach to model the channel success probability is inspired by \cite{pickhardt2021security}. If it is known that the (unknown) liquidity balance $x$ satisfies $\vv{LB}_e \leq x < \vv{UB}_e$, under a uniform probability assumption, it holds that $P(x=b) = \frac{1}{\vv{UB}_e-\vv{LB}_e}$ for any $b$ with $\vv{LB}_e \leq b < \vv{UB}_e$, which implies that we can model the probability that a payment of size $\vv{amt}_e$ through channel $e$ fails as $P(x < \vv{amt}_e) = \sum_{b = \vv{LB}_e}^{\vv{amt}_e-1} P(x=b) = \frac{\vv{amt}_e-\vv{LB}_e}{\vv{UB}_e-\vv{LB}_e}$. Therefore, the success probability can be expressed as the conditional probability 
\begin{equation} \label{eq:conditional:Pe}
\begin{split}
P_e &= P(x \geq \vv{amt}_e \mid \vv{LB}_e \leq x < \vv{UB}_e) \\
    &= 1 - P(x < \vv{amt}_e \mid \vv{LB}_e \leq x < \vv{UB}_e) = \frac{\vv{UB}_e - \vv{amt}_e}{\vv{UB}_e - \vv{LB}_e}.
\end{split}
\end{equation}
CLN and LDK use this approach by assigning default values to the upper and lower bounds of liquidity in a channel based on its capacity. Additionally, in LDK, these bounds are adjusted based on the outcomes of many payment attempts. However, the adjusted bounds gradually revert to their default values over time.

While the previous approaches are based on the assumption of uniform liquidity distribution, assuming a bimodal liquidity distribution is more practical, as the channel liquidity is often unbalanced due to unidirectional flows from the strong sources to strong recipients. This results in liquidity being concentrated either on the local or the remote side \cite{lightningBlazingTrails}. Therefore, the success probability estimation can be based on the assumption of a bimodal distribution \cite{githubPathfindingTake} for the available channel liquidity with density function proportional to 
\begin{equation} \label{eq:exponentialpdf:bimodal}
P(x) \sim e^{\frac{-x}{\vv{s}}} + e^{\frac{x-\vv{cap}_e}{\vv{s}}}
\end{equation}
within the interval $[0, \vv{cap}_e]$.
Here, $\vv{s}$ is a liquidity broadening scale that describes how quickly the bimodal distribution drops off from the extreme balance states of $x = 0$ and $x=\vv{cap}_e$, respectively. A small value of $\vv{s}$, relative to channel capacities, represents a strongly bimodal channel balance distribution. In contrast, a large value of $\vv{s}$ indicates that the channel balances are more uniformly distributed. 

Following this assumption, an unconditional probability estimate for $P_e$ can be derived without requiring prior knowledge, based solely on the overall channel liquidity distribution. A conditional probability estimate, on the other hand, can be obtained based on historical payment attempts, considering the previous successes and failures for a specific channel across multiple attempts. LND's $Bimodal$ mode follows this approach to calculate the channel success probability, where the knowledge of previous payment attempts' successes and failures decays over time. We explain the implementation in \Cref{subsec:LND}.

A bimodal distribution for channel liquidity can alternatively be modeled using the quadratic probability density function (PDF) \cite{githubGitHubLightningdevkitrustlightning} that is proportional to
\[
P(x) \sim \Big(\vv{x - \frac{cap_e}{2}}\Big)^2
\]
within the interval $[0, \vv{cap}_e]$. The quadratic shape of this PDF has modes at $x=0$ and $x = \vv{cap}_e$, but decays only quadratically\footnote{The documentation \cite{githubLDKbimodalscoring} erroneously claims that this PDF is exponential, however, it is a quadratic function.} towards its minimum at $x = \vv{cap}_e/2$, whereas the PDF of \eqref{eq:exponentialpdf:bimodal} decays exponentially. Under this assumption, the channel success probability $P_e$ is determined as the ratio of the integral of the PDF from the amount to the maximum liquidity, to the integral from the minimum to the maximum liquidity. The $Bimodal$ mode in LDK adopts this approach, and its implementation is detailed in \Cref{subsec:LDK_weight}.

\subsection{Attempt Cost/Failure Cost, Hop Cost} \label{ele:attemptcost}
The notions of \emph{attempt, failure or hop costs} play similar roles in the edge weight function \eqref{eq:edgeweight:def} as the fees described in \Cref{sec:fees}. Unlike fees, however, these costs are not proportional to the payment amount $\vv{amt}_e$ and are referred to as virtual costs. They consist of two components: fixed (constant) virtual cost and proportional virtual cost.

A tunable virtual cost $c^i$, representing the cost of a failed payment attempt \cite{githubGitHubLightningnetworklnd}, can be multiplied to the edge-specific success probability term as introduced in \Cref{ele:edgeprob} to balance the trade-off between maximizing success probability and minimizing fees. It is important to choose this cost wisely, as it is ensures payment success through a previously successful path, rather than attempting an untested but potentially less expensive route. 

This virtual cost is called \emph{attempt cost} in LND, where it is multiplied to the path success probability as shown in \eqref{eq:cost:inversePp}, and is referred to as \emph{failure cost} in Eclair, where it is multiplied to the channel success probability.
\subsection{Channel Age}
Each payment channel $e$ has been funded by a certain funding transaction, which had been included in a specific Bitcoin blockchain block. Thus, the \emph{age} of a channel is publicly readily available by computing the difference of the block height of the funding transaction and the current block height. Within pathfinding, the channel age can potentially give an indication on the reliability of the channel for the purpose of payment routing, as it can be argued that older channels are more likely to be suitable for payment routing purposes, as non-performing channels would have been closed by either of the two nodes over a long period of time.

The channel age is currently used as a part of the channel weight definition in Eclair, cf. \Cref{section:eclair}. 

\subsection{HTLC Minimum}
A relevant channel-specific parameter for Lightning pathfinding, which is communicated by LN nodes via gossip, is the \emph{HTLC minimum} parameter $\vv{HtlcMin}_{e}$ or \texttt{htlc\_minimum\_msat}, the minimal payment amount $\vv{amt}_e$ that will be forwarded in a hashed timelock contract (HTLC) routed by the outgoing node of channel $e$ \cite{githubBaseRustyrussell}, \cite[Chapter 10]{antonopoulos2021mastering}.  If the payment amount fails to meet the HTLC minimum of a channel, the sender might need to consider an alternative channel or pay an additional amount to route through the same channel. Consequently, in pathfinding, the sender aims not only to minimize fees associated with routing the payment but also to minimize the extra amount needed to meet the HTLC minimums. LDK uses the HTLC minimum in its weight function to penalize a channel with high HTLC minimum relative to its fee. Another way to avoid any payment failures due to not exceeding the HTLC minimum is to exclude channels with HTLC minimum above the intended payment amount entirely from the LN graph $G$ for the purpose of pathfinding.

\subsection{HTLC Maximum}
A complimentary, but different role plays the \emph{HTLC maximum} parameter $\vv{HtlcMax}_{e}$ or \texttt{htlc\_maximum\_msat}, which indicates an upper bound on the payment amount that an outgoing node in a channel is willing to forward \cite{githubBaseRustyrussell}, \cite[Chapter 10]{antonopoulos2021mastering}. In most channels of the current LN graph, this parameter is chosen to coincide with the capacity $\vv{cap}_e$ \cite{PickhardtValves2022}. If $\vv{HtlcMax}_{e}$ is lower than $\vv{cap}_e$, the channel can be further removed from the pathfinding graph $G$ if the target payment amount exceeds its value. To the best of our knowledge, only LDK uses HTLC maximum information in its weight function by penalizing channels with $\vv{HtlcMax}_{e}$ larger than half of the channel capacity, with the intention to penalize channel probing. 

On the other hand, it could be reasonable to not penalize, but \emph{prioritize} channels with a $\vv{HtlcMax}_{e}$ setting significantly smaller than their capacity, as this setting could serve as a \emph{valve} which regulates the flow payments to induce a better (e.g., more uniform) liquidity distribution within the channel, as pointed out in \cite{PickhardtValves2022}. However, we are not aware of any LN node pathfinding modules that take this point of view into account via their weight functions so far. The difference between the opposite HTLC maximums $\vv{HtlcMax}_{(u,v)}$ and $\vv{HtlcMax}_{(v,u)}$ could also be taken into account as this might induce a change in the liquidity distribution over time \cite{PickhardtValves2022}.

\section{Pathfinding Algorithms}\label{sec:beyond:shortest:path}
In this section, we discuss variants of shortest path algorithms used in LN clients, specific constraints applied during the path selection, resulting implications on the problem complexity class and the optimality of respective solution paths.

\subsection{Shortest Path Algorithms for Additive Costs}
It is folklore in the Lightning Network developer community \cite[Chapter 12]{antonopoulos2021mastering} that the single-path pathfinding problem can simply and efficiently be solved by using Dijkstra's algorithm \cite{Dijkstra-1959note}, which has a worst-case runtime complexity of $O((|E|+|V|)\log(|V|))$ if its priority queue is implemented as a binary heap \cite{Johnson1972-Shortest,Cormen-2022Introduction}, where $|E|$ is the number of channels and $|V|$ is the number of nodes in the Lightning Network graph. This makes Dijkstra's algorithm (presented in \Cref{algo:dijkstra}) a good choice if applicable, even in the presence of significant future growth of the Lightning Network.\footnote{The LN snapshot from 2022 we used \cite{key} in our simulations has $|E| = 2\cdot 57,773$ directed channels, cf. \Cref{sec:evaluation}.}

However, as we see below, some nuance is required, as not all costs defined by LN clients enable such an algorithm to find a path $p$ that minimizes the cost over all eligible paths, and as additional constraints are typically imposed. 

Given the Lightning Network graph $G = (V, E)$, many LN clients use Dijkstra's algorithm \textsc{Dijkstra}$(G, \operatorname{weight}_{a}, s)$ (\Cref{algo:dijkstra}), to find the \emph{cheapest} path from source $s \in V$ node (which is, technically, chosen to be the \emph{recipient} node $r$ of a Lightning payment) to all nodes $u \in V$ in the network (and in particular, to the \emph{sender} of the LN payment) with respect to an additive total path cost $c(p)$ as in \eqref{eq:additive:pathcost}, where $\operatorname{weight}_{a}: E \to \R_{\geq 0}$ is a non-negative channel-specific weight function. On a high level, within the algorithms, distances and predecessors for each vertex $u \in V$ are first initialized. A set $S$ of vertices, whose final shortest-path weights from $s$ have been determined, is maintained. $Q$ is a min-priority queue of vertices, keyed by the total path weight $c_a(p)$ as defined in \eqref{eq:additive:pathcost}. Each vertex $u \in V$ is inserted into the priority queue $Q$, with its priority determined by its initial distance. 

While there exist vertices in $Q$ to process, the vertex $u$ with the smallest distance estimate $u.c_a(p)$ is extracted from $Q$ and the shortest path to $u$ is considered final and added to $S$. For each neighbor $v$ of the vertex $u$, the algorithm attempts to relax the edge $(u, v)$, i.e., the algorithm checks whether the current best-known distance to $v$, $v.c_a(p)$, can be improved by traversing via the edge $(u,v)$. If the distance $\vv{dist}_a$ via $u$, $u.c_a(p) + weight_a(u,v)$, provides a shorter path to $v$, then the predecessor of $v$ can be updated to $u$ and the distance to $v$ can be updated in $Q$, such that $v.c_a(p) = u.c_a(p) + \operatorname{weight}_a(u,v)$.

It can be shown that \textsc{Dijkstra}$(G, \operatorname{weight}_a, s)$ terminates such that the cheapest path to $s$ from each node $u$ in the graph, with respect to the additive path cost 
\begin{equation} \label{eq:additive:pathcost:weighta}
c_a(p) = \sum_{e\in p} \operatorname{weight}_a(e)
\end{equation}
is implicitly found.

However, within LN clients, apart from the total path cost $c_a(p)$ of a path $p$, pathfinding is subject to certain (one or more) \emph{side constraints} that can be expressed as 
\begin{equation} \label{eq:sideconstraint}
\operatorname{x}(p) := \sum_{e \in p} y_e \leq \alpha
\end{equation}
where $\operatorname{x}$ defines the constraint function with non-negative $y_e$, $e \in E$ and $\alpha \geq 0$ an upper bound. The constraints used by LN clients are related to the design elements explored in \Cref{sec:designelements} and try to impose certain payment path properties that are considered desirable, such as a maximum total fee (which, if exceeded by the path fee, makes a path non-desirable by the user) or a minimum path success probability. We summarize existing side constraints used by different LN clients during pathfinding in \Cref{table:side_constraints}. 
\begin{table*}
\centering
  {
\begin{threeparttable}[b]
  \caption{Side constraints used by LN clients during pathfinding}
  \label{table:side_constraints}
  
  \begin{tabular}{lll}
    \toprule
Constraint & & LN Client \\
\midrule
Timelock & $\sum_{e \in p} \timelockdelta \leq \vv{Max\_CLTV\_Limit}$ & LND, LDK, Eclair\tnote{*} \\
Probability & $-\sum_{e \in p} \log (P_e) \leq -\log (Min\_Prob\_Limit)$ & LND, LDK \\ 
Fee & $\sum_{e \in p} \vv{fee}_e \leq \vv{Max\_Fee\_Limit}$ & LND, LDK, Eclair\tnote{*} \\
Path Length & $\sum_{e \in p} 1 \leq \vv{Max\_PathLength\_Limit}$ & LDK, CLN\tnote{*}, Eclair\tnote{*}
\\
\bottomrule
\bottomrule
  \end{tabular}
  \begin{tablenotes}
       \item [*] Constraints are validated after pathfinding.
     \end{tablenotes}
    \end{threeparttable}
    }
\end{table*}

In practice, the adherence of paths to a constraint \eqref{eq:sideconstraint} is achieved by adding the colored lines in \Cref{algo:dijkstra}, which makes sure that any constructed path does not violate \eqref{eq:sideconstraint}. In particular, \Cref{algo:dijkstra} only updates the the distance of $v$ in $Q$ and the predecessor of $v$ for a neighbor node $v$ of $u$ if the new constraint value $\vv{cstval} = u.x(p) + y(u,v)$ does not exceed the bound $\alpha$, i.e., if $ \vv{cstval} \leq \alpha$. While the colored modification in \Cref{algo:dijkstra} only implies the adherence to one constraint \eqref{eq:sideconstraint}, this modification can be analogously extended to multiple constraints $x_1(p) \leq \alpha_1,\ldots,x_m(p) \leq \alpha_m$.

\begin{algorithm}
\caption{Dijkstra's Algorithm \cite{Dijkstra-1959note} (with \colorbox{yellow}{colored} modification if to impose constraint $\operatorname{x}(p) \leq \vv{alpha}$) for cost \eqref{eq:additive:pathcost:weighta}}
\label{algo:dijkstra}
\begin{algorithmic}
\State \textsc{Dijkstra}($G, \operatorname{weight}_a, $\colorbox{yellow}{$y$}, \colorbox{yellow}{$\vv{alpha}$}$, s$)
    \For{each vertex $u \in V$}
        \State $u.c_a(p) \gets +\infty$
        \State $u.\vv{prev} \gets$ \textsc{UNDEFINED}
    \EndFor
    \State $S \gets \emptyset$
    \State $Q \gets \emptyset$
    \State $s.c_a(p) \gets$ $0$
    \State \colorbox{yellow}{$s.x(p) \gets$ $0$}
    \For {each vertex $u \in V$}
        \State \textsc{Insert}(Q, u)
    \EndFor
    \While{$Q \neq \emptyset$}
        \State $u \gets$ \textsc{Extract-Min}($Q$)
        \State $S \gets S \cup \{u\}$
        \For{each vertex $v \in Adj[u]$}
            \State $\vv{dist}_a \gets u.c_a(p) + \operatorname{weight}_a(u,v)$
            \State \colorbox{yellow}{$\vv{cstval} \gets u.x(p) + y(u,v)$}
            \If {$\vv{dist}_a < v.c_a(p)$ \colorbox{yellow}{\& $\vv{cstval} \leq \vv{alpha}$}}
                \State $v.\vv{prev} \gets u$
                \State $v.c_a(p) \gets \vv{dist}_a$
                \State \colorbox{yellow}{$v.x(p) \gets \vv{cstval}$}
                \State \textsc{Decrease-Key}($Q, v, \vv{dist}_a$)
            \EndIf
        \EndFor
    \EndWhile
\end{algorithmic}
\end{algorithm}

The constraint-adhering version \textsc{Dijkstra}$(G, \operatorname{weight}_{a}, y, \vv{alpha}, s)$ of \Cref{algo:dijkstra} can be interpreted as a \emph{greedy algorithm} for solving the so-called \emph{constrained shortest-path problem} (with respect to cost \eqref{eq:additive:pathcost:weighta} and constraint \eqref{eq:sideconstraint}), a problem class that has been studied by a considerable body of literature \cite[Chapter 16]{Ahuja-1993Network}, \cite[Section 8.4]{bertsekas1998network}, \cite{Xiao-2005constrained,Irnich-2005shortest,Pugliese-2013Survey,Bendahi2024constrained}. 

While the conventional shortest path problem can be solved, e.g., by \Cref{algo:dijkstra} in a time that is essentially \emph{linear} in $|E|$ as mentioned above, the constrained shortest-path problem is known to be $\mathcal{NP}$-complete, \cite[Appendix B.4]{Ahuja-1993Network},\cite[Section 8.4]{bertsekas1998network}, which implies that, unless $\mathcal{P} = \mathcal{NP}$, no polynomial-time algorithm exists that can solve the constrained shortest-path problem for all problem instances. Thus, the output of \textsc{Dijkstra}$(G, \operatorname{weight}_{a}, y, \vv{alpha}, s)$ will not be able to optimally solve the problem it is designed for solving in general. 

\subsection{Shortest Path Algorithms for Additive+Multiplicative Costs} \label{sec:shortestpath:addmultcost}
We discussed in \Cref{ele:successprob} that path cost modeling used in the LND LN client implementation amounts to an additive-plus-multiplicative cost function $c(p)$, which was defined in \eqref{eq:mixed:additivemult:cost}. In order to find paths $p$ that minimize such a cost $c(p)$, LND uses a modified Dijkstra-style algorithm, which we outline in \Cref{algo:modified-dijkstra}. In particular, compared to \Cref{algo:dijkstra}, the priority queue is keyed based on the additive-plus-multiplicative cost $c(p)$ instead of on a purely additive cost \eqref{eq:additive:pathcost:weighta}, and the algorithm \textsc{ModDijkstra}$(G, \operatorname{weight}_{a},\operatorname{weight}_{m},c_{\vv{attempt}}, s)$ takes as additional input a multiplicative weight function $\operatorname{weight}_{m}: E \to \R_{\geq 1}$ with outputs larger than or equal $1$, and an initial cost term factor $c_{\vv{attempt}}$. 

Similar to \Cref{algo:dijkstra}, after initialization, each neighboring vertex $v$ of $u \in Q$ is processed. During the relaxation of the edge $(u,v)$, the algorithm computes the total additive weight $\vv{dist}_a$ as $u.c_a(p) + \vv{weight}_a(u,v)$ and the total multiplicative weight $\vv{dist}_m$ as $u.c_m(p)\cdot \vv{weight}_m(u,v)$. The predecessor of $v$ is updated to $u$ and the additive-plus-multiplicative distance $\vv{dist}_a + \vv{dist}_m$ is updated for $v$ in $Q$, provided the distance $\vv{dist}_a + \vv{dist}_m$, is greater than the previous distance to $v$, $v.c_a(p) + v.c_m(p)$.
\begin{algorithm}
\caption{Dijkstra-Style Algorithm Modified for Additive+Multiplicative Cost \eqref{eq:mixed:additivemult:cost} as used in LND (with addition of \textcolor{blue}{blue lines} compared to \Cref{algo:dijkstra} and \colorbox{yellow}{yellow-background} lines if to also impose constraint $\operatorname{x}(p) \leq \vv{alpha}$)}
\label{algo:modified-dijkstra}
\begin{algorithmic}
\State \textsc{ModDijkstra}($G, \operatorname{weight}_a, \textcolor{blue}{\operatorname{weight}_m},\textcolor{blue}{c_{\vv{attempt}}},$\colorbox{yellow}{$y$}, \colorbox{yellow}{$\vv{alpha}$}$,s$)
    \For{each vertex $u \in V$}
        \State $u.c_a(p) \gets +\infty$
        \State \textcolor{blue}{$u.c_m(p) \gets 0$} 
        \State $u.\vv{prev} \gets$ \textsc{UNDEFINED}
    \EndFor
    \State $S \gets \emptyset$
    \State $Q \gets \emptyset$
    \State $s.c_a(p) \gets 0$
    \State \textcolor{blue}{$s.c_m(p) \gets c_{\vv{attempt}}$}
    \State \colorbox{yellow}{$s.x(p) \gets$ $0$}
    \For {each vertex $u \in V$}
        \State \textsc{Insert}(Q, u)
    \EndFor
    \While{$Q \neq \emptyset$}
        \State $u \gets$ \textsc{Extract-Min}($Q$)
        \State $S \gets S \cup \{u\}$
        \For{each vertex $v \in Adj[u]$}
            \State $\vv{dist}_a \gets u.c_a(p) + \operatorname{weight}_a(u,v)$
            \State \textcolor{blue}{$\vv{dist}_m \gets u.c_m(p) \cdot \operatorname{weight}_m(u,v)$}
            \State \colorbox{yellow}{$\vv{cstval} \gets u.x(p) + y(u,v)$}
            \If {$(\vv{dist}_a\textcolor{blue}{+\vv{dist}_m)} < (v.c_a(p)\textcolor{blue}{+v.c_m(p)})$ \colorbox{yellow}{\& $\vv{cstval} \leq \vv{alpha}$}}
                \State $v.\vv{prev} \gets u$
                \State $v.c_a(p) \gets \vv{dist}_a$
                \State \textcolor{blue}{$v.c_m(p) \gets \vv{dist}_m$}
                \State \colorbox{yellow}{$v.x(p) \gets \vv{cstval}$}
                \State \textsc{Decrease-Key}($Q, v, \vv{dist}_a \textcolor{blue}{+ \vv{dist}_m}$)
            \EndIf
        \EndFor
    \EndWhile
\end{algorithmic}
\end{algorithm}

It turns out that, even without the imposition of any side constraints \eqref{eq:sideconstraint} (no colored modification in \Cref{algo:modified-dijkstra}), \Cref{algo:modified-dijkstra} is not guaranteed to find the optimal path due to the multiplicative weight used in $c(p)$ of \eqref{eq:mixed:additivemult:cost}. 

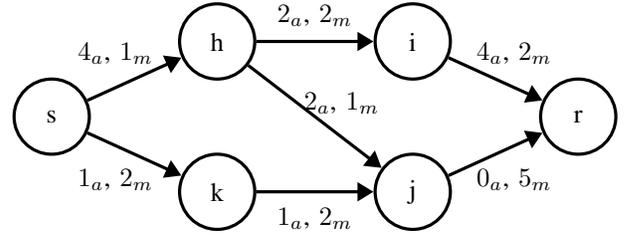
\begin{figure}[h!]
    \centering
    \begin{tikzpicture}
    \node[mynode](n1) at (-3, 1){s};
    \node[mynode](n2) at (-0.8, 2){h};
    \node[mynode](n3) at (1.8, 2){i};
    \node[mynode](n4) at (4, 1){r};
    \node[mynode](n5) at (-0.8, 0){k};
    \node[mynode](n6) at (1.8, 0){j};
    
    \draw[myarrow](n1)--node[above=5pt, pos=0.3]{$4_a$, $1_m$}(n2);
    \draw[myarrow](n2)--node[above=2pt]{$2_a$, $2_m$}(n3);
    \draw[myarrow](n3)--node[above=5pt, pos=0.7]{$4_a$, $2_m$}(n4);
    \draw[myarrow](n1)--node[below=5pt, pos=0.3]{$1_a$, $2_m$}(n5);
    \draw[myarrow](n5)--node[below=2pt]{$1_a$, $2_m$}(n6);
    \draw[myarrow](n6)--node[below=5pt, pos=0.7]{$0_a$, $5_m$}(n4);
    \draw[myarrow](n2)--node[above=5pt, pos=0.7]{$2_a$, $1_m$}(n6);
    \end{tikzpicture}
\caption{Counterexample graph $G$ for suboptimality of \Cref{algo:modified-dijkstra} for pathfinding with respect to additive-plus-multiplicative cost \eqref{eq:mixed:additivemult:cost}.}
\label{fig:dijkstra_counter_example}
\end{figure}

To illustrate the suboptimality of \textsc{ModDijkstra}$(G, \operatorname{weight}_{a},\operatorname{weight}_{m},c_{\vv{attempt}}, s)$, consider a directed graph $G=(V, E)$ as shown in \Cref{fig:dijkstra_counter_example}. This graph consists of six nodes $V=\{r, s, h, i, j, k\}$ and seven edges $E=\{(s,h), (s,k), (h,i), (h,j), (k, j), (i,r), (j,r)\}$. Each edge $e \in E$ comprises of an additive weight $\operatorname{weight_a(e)}$ and a multiplicative weight $\operatorname{weight_m(e)}$, denoted with subscript $a$ and with subscript $m$ in \Cref{fig:dijkstra_counter_example}, respectively. To calculate the total path cost, we use \eqref{eq:mixed:additivemult:cost}, where the virtual attempt cost $c_{\text{attempt}}$ is taken as $1$. For simplicity, we do not impose side constraints such as \eqref{eq:sideconstraint} in this example--additional side constraints make the problem even harder, and suboptimality is easier to show.

To find a path from source $s$ to recipient $r$ using \Cref{algo:modified-dijkstra}, the algorithm first initializes the additive weight $c_a(p)$ and the multiplicative weight $c_m(p)$ for each vertex to $\infty$ and $0$, respectively, then set $S$, and the priority queue $Q$. Starting from source $s$, we relax the adjacent nodes, $h$ and $k$, of $s$. From $s$, the distance to $h$ is calculated as $h.c_a(p) + h.c_m(p) = 4 + 1 = 5$. Similarly, $k.c_a(p) + k.c_m(p) = 1 + 2 = 3$. The priority queue is keyed based on the distances for subsequent relaxations as $Q = \{(k, 3), (h, 5)\}$. Now, $k$ is removed from the priority queue and its neighbor $j$ is relaxed. The distance to $j$ is calculated as $1+1+2 \cdot 2 = 6$. Since the distance to $j$ from $s$, via $k$, is greater than the distance to $h$ from $s$, node $j$ will be added to the end of $Q$ as $\{(h, 5), (j, 6)\}$. As a next step, we therefore remove $h$ from the queue, and the distance to its neighbor $i$ is obtained as $4+2+2 = 8$, and we get the distance to $j$ as $4+2+1 = 7$. However, the already known distance to $j$ via $k$ is $6$ which is shorter than the distance via $h$. Consequently, the distance to $j$ in $Q$ will not be updated based on the distance via $h$. With $Q = \{(j,6), (i,8)\}$, node $j$ will be removed from the queue, and the edge to its neighbor $r$ is relaxed, giving the distance to $r$ as $1+1+2\cdot 2\cdot 5 = 22$. Then, node $i$ will be removed from the queue and its edge to $r$ is relaxed, yielding the distance to $r$ as $4+2+4+1 \cdot 2\cdot 2 = 14$. Hence, the algorithm terminates and returns the path $s\to h\to i\to r$ with the cost of $r.c_a(p) + r.c_m(p) = 14$.

However, this is not the optimal path, as the cost of the path $s\to h\to j\to r$ is $4+2+1\cdot 1\cdot 5 = 11$, which is cheaper than the path returned by \Cref{algo:modified-dijkstra}. The algorithm ignores this path during relaxation, as illustrated, due to the nature of the additive-plus-multiplicative cost as in \eqref{eq:mixed:additivemult:cost}, which leads the algorithm to greedily select the predecessor based on the current weight, thereby excluding potentially less expensive routes. Via this counterexample, we conclude that, while the algorithm might often behave reasonably, the modified Dijkstra-style algorithm \textsc{ModDijkstra}$(G, \operatorname{weight}_{a},\operatorname{weight}_{m},c_{\vv{attempt}}, s)$ used in LND cannot \emph{guarantee} to return an optimal path with respect to its modeled cost function \eqref{eq:mixed:additivemult:cost}.

While it is possible to tune the value of $c_{\text{attempt}}$ to give more weight to the multiplicative weight $c_m(p)$ in \eqref{eq:mixed:additivemult:cost} to fix the behavior for this example, a counterexample for optimality can be found for any choice of $c_{\text{attempt}}$.

\subsection{$K$-Shortest Path Algorithms}
Beyond Dijkstra-type algorithms to find a single shortest path between sender and receiver nodes of a payment with respect to a fixed path cost function as presented in \Cref{algo:dijkstra} and \Cref{algo:modified-dijkstra}, there are also LN clients that return a \emph{selection} of candidate paths during pathfinding.

For this purpose, in particular, Yen's $K$-shortest path algorithm \cite{yen1971finding} is used. This algorithm uses a shortest path algorithm (such as \Cref{algo:dijkstra}) as a subroutine, but returns a list of $K$ ``best'` paths with respect to the imposed cost function. If Dijkstra's algorithm is used as a subroutine, and if no constraints \eqref{eq:sideconstraint} are imposed, the worst-case time complexity of the algorithm is $O(K |V| (|E|+|V|)\log(|V|))$ (for Dijkstra implemented with binary heap) \cite[Section 6.3]{Bouillet2007path}. This complexity might limit the scalability of the approach in the case of further growth of the Lightning Network. Yen's $K$-shortest path algorithm is currently used only in the Eclair LN client.

\section{Implementation of Pathfinding in Lightning Node Clients} \label{sec:client:weights}
In this section, we detail the single-path pathfinding strategies employed by the four popular Lightning node clients LND, CLN, LDK, and Eclair, with a particular focus on a breakdown of the defined cost functions/ edge weights, including relevant constants and default values. Each of these Lightning node clients finds a payment path using one of the algorithms discussed in \Cref{sec:beyond:shortest:path}. LND utilizes the modified Dijkstra-style algorithm \Cref{algo:modified-dijkstra}, CLN, and LDK use \Cref{algo:dijkstra}, whereas Eclair utilizes Yen’s $\vv{K}$-shortest path algorithm. Each of them uses significantly different weight functions $\operatorname{weight}: E \to \R_{\geq 0}$ with different trade-offs. To accurately compute the fee $\vv{fee}_e$ associated with using a channel between nodes, all these clients perform the path search backwards from the receiver node to the source node.
\subsection{Lightning Network Daemon (LND)}\label{subsec:LND}
In LND, the weight function considers various design elements mentioned in \Cref{sec:designelements}, including the payment amount and fee for using the channel $e$ as well as the CLTV delta $\timelockdelta$ and a penalty term related to the estimated success probability. Specifically, the weight function is defined as 
\begin{equation}\label{weight:lnd}
\begin{aligned}
    \vv{weight(e,p)} &= \vv{fee}_e + \vv{amt}_e \cdot \timelockdelta \cdot \, \vv{riskfactor} \\
    &\quad + \frac{\vv{penalty}_e}{\vv{P}_{p}},
\end{aligned}
\end{equation}
where $\vv{fee}_e$ and $\timelockdelta$ are specific to the channel and are obtained through the channel update messages \cite{githubGitHubLightningbolts} within the network, and $\vv{amt}_e$ is the payment amount through channel $e$. In LND, the influence of the $\timelockdelta$ is controlled by scaling it using the constant $\vv{riskfactor}$, which defaults to a value of $15 \cdot 10^{-9}$. In \eqref{weight:lnd}, the edge penalty $\vv{penalty}_e$ is defined as 
\begin{equation}\label{penalty:lnd}
\begin{aligned}
    \vv{penalty}_e = \vv{attemptcost}_e \cdot \frac{1}{0.5 - \frac{0.9 \cdot \vv{timepref}}{2}}-1,
\end{aligned}    
\end{equation}
where $\vv{timepref}$ is a time preference constant for the payment. It is set to $-1$ to prioritize optimizing for fees only, set to $1$ to prioritize optimizing for reliability only, or set to a value in between for a balanced approach. By default, its value is $0$. Through \eqref{penalty:lnd}, LND's weight function also incorporates the design element \emph{attempt cost} (\Cref{ele:attemptcost}) which is defined as
\begin{equation}\label{attemptcost:lnd}
\begin{aligned}
    \vv{attemptcost}_e &= \vv{BaseAttemptCost} \\
    &\quad + \vv{amt}_e \cdot \vv{AttemptCostRate}.
\end{aligned}
\end{equation}
with constants $\vv{BaseAttemptCost}$ and $\vv{AttemptCostRate}$. This is a predetermined virtual cost that signifies the cost of a failed payment attempt. The default values of $\vv{BaseAttemptCost}$ and $\vv{AttemptCostRate}$ are $100$ msat and $1000$ msat, respectively.

As can be seen in \eqref{weight:lnd}, LND multiplies $\vv{penalty}_e$ from \eqref{penalty:lnd} with the inverse of the success probability $\vv{P_p}$ of the path $p$, which is a path-dependent cost that prevents the applicability of a standard shortest path algorithm such as \Cref{algo:dijkstra} (cf. \Cref{sec:shortestpath:addmultcost}). LND assumes independence of the channel success probabilities, which leads via $P_p = \prod_{e \in p}P_{e}$ (cf. \eqref{lnd:probpath}) to an additive-plus-multiplicative total path cost $c(p)$. Therefore, LND uses the modified Dijkstra algorithm as outlined in \Cref{algo:modified-dijkstra} in its pathfinding. 

LND keeps records of past failed payment attempts within a channel, allowing for the tracking of the time elapsed since the last failure. This information is used in the estimation of channel success probabilities $P_e$, underscoring LND's emphasis on prioritizing reliable routing paths. LND uses two distinct methods for calculating the channel-wise success probability $\vv{P_e}$ of a channel forwarding a payment: $Apriori$ and $Bimodal$. 

In $Apriori$, the success probability is influenced by the weight of the last failure. Initially, when a failure occurs, the success probability $P_e$ drops to $0$. However, over time, based on the defined $\vv{PenaltyHalfLife}$ period, the success probability gradually recovers to its node probability ($\vv{nodeProb}_e$) via the formula
\begin{equation} \label{prob:apriori:lnd}
\begin{aligned}
    P_e &= \vv{nodeProb}_e \cdot \left(1-\frac{1}{2^{\frac{\vv{timeSinceLastFailure}_e}{\vv{PenaltyHalfLife}}}}\right).
\end{aligned}
\end{equation}

In the above equation, $\vv{nodeProb}_e$ is given by,
\begin{equation} \label{prob:nodeprob:lnd}
    \vv{nodeProb}_e = P_{apriori} \cdot \left(1-\frac{0.5}{1+e^{\frac{\vv{c}_o \cdot \vv{cap}_e - \vv{amt}_e}{\vv{s}_o \cdot \vv{cap}_e}}}\right).
\end{equation}

The assumption made by LND, in this case, is that each hop has $\vv{P}_{apriori}$ (base success likelihood) of 0.6. The values of $\vv{c}_o$ and $\vv{s}_o$ in \eqref{prob:nodeprob:lnd}, which represent the fraction of channel capacity $\vv{cap}_e$ at which probability reweighing becomes noticeable and the rate at which probability adjustments occur \cite{lightningClearingPaths}, respectively, are set to 0.9999 and 0.025. The preset $\vv{PenaltyHalfLife}$ is 1 hour. 

In the $Bimodal$ estimation of $P_e$, LND assumes a bimodal distribution $\vv{P(x)} \sim e^{\frac{\vv{-x}}{\vv{s}}} + e^{\frac{\vv{x-cap_e}}{\vv{s}}}$ as explained in \Cref{ele:edgeprob}. The liquidity broadening scale $\vv{s}$ gets the default value of ${3\cdot 10^{5}\,\vv{sats}}$.

In this context, the success probability for a payment amount is determined by integrating over the prior distribution, $\vv{P(x)}$. However, if information regarding previous success and failure amounts is available, the prior distribution $\vv{P(x)}$ is being adjusted. The posterior distribution is then derived by re-normalizing the prior distribution such that 
\begin{equation} \label{prob1:bimodal:lnd}
\begin{aligned}
    P_e = \frac{\int_{\vv{amt}_e}^{\vv{fa}_e} \vv{P(x)} \, \vv{dx}}{\int_{\vv{sa}_e}^{\vv{fa}_e} \vv{P(x) \, dx}}.
\end{aligned}
\end{equation}
Upon solving the integral, we obtain the formula 
\begin{equation} \label{prob2:bimodal:lnd}
   \begin{aligned}
   P_e &= \frac{e^{\frac{(\vv{fa}_e - \vv{cap}_e)}{\vv{s}}} - e^{\frac{-\vv{fa}_e}{\vv{s}}} - e^{\frac{(\vv{amt}_e - \vv{cap}_e)}{\vv{s}}} + e^{\frac{-\vv{amt}_e}{\vv{s}}}}{e^{\frac{(\vv{fa}_e - \vv{cap}_e)}{\vv{s}}} - e^{\frac{-\vv{fa}_e}{\vv{s}}} - e^{\frac{(\vv{sa}_e - \vv{cap}_e)}{\vv{s}}} + e^{\frac{-\vv{sa}_e}{\vv{s}}}}
   \end{aligned}
\end{equation}
for the channel-wise success probability $P_e$,
 where $\vv{sa}_e$ and $\vv{fa}_e$ represent the unsettled success and failure amounts, respectively and $\vv{cap}_e$ is the channel capacity. 

From \eqref{prob2:bimodal:lnd}, it can be noted that the $Bimodal$ estimator incorporates information previous success and failure amounts, unlike the $Apriori$ estimator \eqref{prob:apriori:lnd}, which does not incorporate such information.

As detailed in \Cref{table:side_constraints}, LND enforces side constraints during pathfinding in \Cref{algo:modified-dijkstra} to ensure compliance with specific limits. These include $\sum_{e \in p} \timelockdelta \leq \vv{Max\_CLTV\_Limit}$, $\prod_{e \in p} P_e \geq 0.01$, and $\sum_{e \in p} \vv{fee}_e \leq \vv{Max\_Fee\_Limit}$. Here, $\vv{Max\_CLTV\_Limit}$ and $\vv{Max\_Fee\_Limit}$ represent user-defined maximum allowable CLTV and fee limits, respectively.

Overall, the pathfinding strategies used by LND are founded on the principle of assessing channels with high reliability, characterized by low fees and short timelocks. This rationale is also reflected in the weight function's consideration of fees, timelock, and success probability.
\subsection{Core Lightning (CLN)}
Design elements considered in the weight function of CLN are success probability, fees and timelock, leading to the weight function
\begin{equation}\label{weight:cln}
\begin{aligned}
    \vv{weight}(e) &= \left(\vv{fee}_e +\frac{\vv{amt}_e \cdot \timelockdelta \cdot \, \vv{riskfactor}}{\vv{BlockPerYear} \cdot 100} + 1\right) \\
    &\quad \cdot \left(\vv{capacityBias}_e + 1\right)
\end{aligned}
\end{equation}
for the channel $e$. Similar to its application in LND as observed in \eqref{weight:lnd} with $\timelockdelta$, a constant $\vv{riskfactor}$ is used here in conjunction with $\timelockdelta$ to regulate its impact. The default value for $\vv{riskfactor}$ is $10$. The denominator in \eqref{weight:cln} is motivated from a yearly expense incurred by funds locked within a channel, and uses the constant $\vv{BlockPerYear} =52,596$ due to the expected number of blocks per year in the Bitcoin network. 

Additionally, CLN introduces $\vv{capacityBias}_{e}$ to penalize channels where a substantial portion of the channel's total capacity is being utilized. The formulation of $\vv{capacityBias}_e$ can be derived from a modeling of the path success probability $P_e$ with a negative logarithm penalization (cf. \eqref{eq:log_prob}) and the assumption of a uniform liquidity distribution similar to  \eqref{eq:uniform:modeling}, and is given by
\begin{equation}\label{capbias:cln}
\begin{aligned}
\vv{capacityBias}_e = -\log\left(\frac{\vv{cap}_e+1 - \vv{amt}_e}{\vv{cap}_e+1}\right),
\end{aligned}
\end{equation}
where $\vv{amt}_e$ is the channel payment amount and $\vv{cap}_e$ is the channel capacity.

Overall, the breakdown of \eqref{weight:cln} illustrates a notable emphasis of CLN's pathfinding on $\timelockdelta$ through the selection of higher value for the $\vv{riskfactor}$. Consequently, CLN is inclined to favor paths characterized by lower timelock. CLN uses this weight function \eqref{weight:cln} in \Cref{algo:dijkstra} and constraint $\sum_{e \in p} 1 \leq 10$ is validated after pathfinding. 

\subsection{Lightning Development Kit (LDK)} \label{subsec:LDK_weight}
LDK's pathfinding incorporates the design elements fees, success probability, HTLC minimum, HTLC maximum into its weight function. The resulting channel weight function of LDK can be defined via
\begin{equation}\label{weight:ldk}
\begin{aligned}
    \vv{weight(e)} = \max(\vv{fee}_e,\ \vv{pathHtlcMin}_e) + \vv{penalty}_e,
\end{aligned}
\end{equation}
where $\vv{fee}_e$ is the total fee \eqref{one}, $\vv{penalty}_e$ a multifaceted term discussed below, and $\vv{pathHtlcMin}_e$ a prospective fee cost of related to a payment of size of the minimum HTLC. In particular, payments must meet the minimum HTLC limit set by each channel that is used as described in \Cref{sec:designelements}. Therefore, to utilize the channel by meeting the HTLC minimum, the payment source might have to pay an additional amount along with the actual payment amount, leading to the term 
\begin{equation}\label{pathhtlc:ldk}
\begin{aligned}
    \vv{pathHtlcMin}_e &= \vv{HtlcMin}_e \cdot(1 + \vv{f}_e) + \vv{b}_e,
\end{aligned}
\end{equation}
where $\vv{HtlcMin}_{e}$ is the channel-specific HTLC minimum parameter, $\vv{b}_e$ the base fee and $\vv{f}_e$ the proportional fee charged by the channel, which are all obtained from the channel update message \cite{githubGitHubLightningbolts}.

The term $\vv{penalty}{_e}$ in the channel weight is the sum of four terms related to a base penalty, an anti-probing penalty, a liquidity penalty and a penalty based on a historical estimate, such that
\begin{equation}\label{penalty:ldk}
\begin{aligned}
    \vv{penalty}_e &= \vv{basePenalty}_e + \vv{antiProbingPenalty}_e \\
    &\quad + \vv{liquidityPenalty}_e + \vv{historicPenalty}_e,
\end{aligned}
\end{equation}
where the base penalty
\begin{equation}\label{base:penalty:ldk}
    \vv{basePenalty}_e = \vv{penaltyBase} + \frac{\vv{baseMultiplier} \cdot \vv{amt}_e}{2^{30}}
\end{equation}
consists of a fixed penalty $\vv{penaltyBase}$ with a default value of $500$ msat and the constant  $\vv{baseMultiplier}$, which is multiplied to the  payment amount $\vv{amt}_e$ of the channel and whose default value is $\vv{baseMultiplier} = 8,192\,\vv{msat}$.

The anti-probing penalty in \eqref{penalty:ldk} is set to 250 msat is applied if a channel's HTLC maximum  ($\vv{HtlcMax}_{e}$) is equal to or greater than half of the channel's capacity $\vv{cap}_e$, and set to $0$ otherwise, in order to prefer channels with smaller $\vv{HtlcMax}_{e}$, i.e.,
\begin{equation}\label{antiprobing:ldk}
\begin{aligned}
    \vv{antiProbingPenalty}_e = \begin{cases}
          250, & \text{if } \vv{HtlcMax}_e \geq \frac{\vv{cap}_e}{2}, \\
          0, & \text{otherwise}.
        \end{cases}
\end{aligned}
\end{equation}
The liquidity penalty, which incorporates success probability modeling, is defined as  \begin{equation}\label{liquidity:penalty:ldk}
\begin{aligned}
    \vv{liquidityPenalty}_e = &-\log_{10}(P_e) 
    \\
    &\cdot (\vv{LM} + \vv{amtMultiplier}_e),
 \end{aligned}
\end{equation}
where $\vv{LM} = 30,000\,\vv{msat}$ is a constant multiplier, $\vv{amtMultiplier}_{e}$ is defined below in \eqref{amt:mul:ldk}, and $P_e$ is a channel success probability estimate (see discussion of $P_e$ at the end of this subsection), based on the latest estimate of channel liquidity.

$\vv{historicPenalty}{_e}$ is similar to $\vv{liquidityPenalty}{_e}$. However, in this case, instead of solely relying on the latest estimate of the liquidity available in the channel, the calculation of the success probability is based on the historical estimated liquidity available in the channel. 
A constant multiplier $\vv{HM}$ is also utilized alongside the negative logarithm of the channel's success probability $\vv{hP_e}$ for the payment. The calculation of $\vv{hP_e}$ is also performed using $P_e$, as discussed at the end of this section, based on historical estimates of the channel's available liquidity, rather than the most recent liquidity estimates, as in the case of $P_e$ in $\vv{liquidityPenalty}_e$. 
$\vv{historicPenalty}_e$, with the constant multiplier $\vv{HM} = 10,000\,\vv{msat}$, is defined as
\begin{equation}\label{hist:penalty:ldk}
\begin{aligned}
   \vv{historicPenalty}_e &= -\log_{10}\vv{(hP_e)} \\
   &\quad \cdot (\vv{HM} + \vv{amtMultiplier}_e).
\end{aligned}
\end{equation}

$\vv{amtMultiplier}_{e}$ in \eqref{liquidity:penalty:ldk} and \eqref{hist:penalty:ldk} is given by,
\begin{equation}\label{amt:mul:ldk}
    \vv{amtMultiplier}_e = \frac{\vv{Multiplier}\cdot \vv{amt}_e}{2^{20}},
\end{equation}
where $\vv{Multiplier}=\vv{liquidityPenaltyMultiplier} = 192\,\vv{msat}$ for $\vv{liquidityPenalty}{_e}$ and  $\vv{Multiplier}=\vv{historicPenaltyMultiplier}  = 64\,\vv{msat}$ for $\vv{historicPenalty}{_e}$, respectively.

Within LDK, there are two different options for estimating the channel success probability $P_e$: the first assumes a uniform channel liquidity distribution, and the second assumes a bimodal liquidity distribution. Under the uniform liquidity assumption, $P_e$
is calculated using upper and lower liquidity bounds $\vv{UB}_e$ and $\vv{LB}_e$ that are determined using the past unsuccessful and successful payments based on the formula
\begin{equation}\label{prob:ldk}
   P_e = \frac{\vv{UB}_e - \vv{amt}_e}{\vv{UB}_e - \vv{LB}_e},
\end{equation} 
see also \eqref{eq:conditional:Pe}.

As a second option, LDK can assume a bimodal liquidity distribution given by $P(x) \sim \Big(\vv{x - \frac{cap_e}{2}}\Big)^2$, where channel liquidities are concentrated at the edges. Under this assumption, $P_e$ is computed as the integral under the curve from the payment amount in the channel to the channel's upper bound, divided by the integral under the curve from the lower bound to the upper bound of the channel liquidity. This is expressed as:
\begin{equation} \label{prob:ldk_bimodal_integral}
\begin{aligned}
    P_e &= \frac{\int_{\vv{amt}_e}^{\vv{UB}_e} \vv{P(x)} \, \vv{dx}}{\int_{\vv{LB}_e}^{\vv{UB}_e} \vv{P(x) \, dx}}
\end{aligned}
\end{equation}
Therefore, we obtain
\begin{equation} \label{prob:ldk_bimodal}
   \begin{aligned}
   P_e &= \frac{\left((\vv{UB}_e - \frac{\vv{cap}_e}{2})^3 - (\vv{amt}_e - \frac{\vv{cap}_e}{2})^3 \right) \cdot 64 \cdot 1024^3 + 1}
   {\left((\vv{UB}_e - \frac{\vv{cap}_e}{2})^3 - (\vv{LB}_e - \frac{\vv{cap}_e}{2})^3\right) \cdot 64 \cdot 1024^3 + 1}.
   \end{aligned}
\end{equation}

As the weight function of LDK \eqref{weight:ldk} considers the maximum value between fee and HTLC minimum, selecting a channel with the minimum fee may not always be possible. In scenarios where two paths have the same fees but different HTLC minimums, the path with higher HTLC minimum may be chosen. This can compel the sender to pay additional fees to meet the HTLC minimum, especially when the payment amount is small, potentially resulting in higher overall fees for the sender.

LDK uses the weight function \eqref{weight:ldk} in \Cref{algo:dijkstra}. Similar to LND, LDK also imposes side constraints during pathfinding, as outlined in \Cref{table:side_constraints}. These include: the timelock constraint $\sum_{e \in p} \timelockdelta \leq 1008$, the probability constraint $-\sum_{e \in p} \log (P_e) \leq -\log (0.01)$, the fee constraint $\sum_{e \in p} \vv{fee}_e \leq \vv{(1\%\ of\  payment\ amount + 50\ sats)}$, and the path length constraint $\sum_{e \in p} 1 \leq \vv{19}$.

\subsection{Eclair} \label{section:eclair}
In Eclair, weight computation involves three distinct cases, which are discussed in detail below.
\subsubsection{Case 1: Heuristics ratios}

In this case, weight of an edge is based on the design elements including fee, channel age, CLTV delta, capacity and a virtual hop cost, such that
\begin{equation}\label{weight:eclair}
    \vv{weight(e)} = (\vv{fee}_e + \vv{HopCost}_e) \cdot \vv{factor}_e,
\end{equation}
where $\vv{factor}_e$ is given by 
\begin{equation}
\begin{aligned}
\label{factor:eclair}
    \vv{factor}_e & = \vv{basefactor} + (n_{cltv_e} \cdot \vv{CLTVfactor}) \\
    & \quad + (\vv{n}_{\vv{age}_e} \cdot \vv{agefactor}) + ((1 - \vv{n}_\vv{{cap}_e}) \cdot \vv{capfactor}).
\end{aligned}
\end{equation}
$\vv{factor}{_e}$ includes heuristics ratios and the normalized values. Eclair utilizes default ratios 0.35, 0, 0.5 and 0.15 for $\vv{agefactor}$, $\vv{basefactor}$, $\vv{capfactor}$ and $\vv{CLTVfactor}$ respectively.
Eclair computes the normalized values of $\timelockdelta$ ($\vv{n}{_{\vv{cltv}_e}}$), channel age ($\vv{n}{_\vv{age_e}}$) and capacity ($\vv{n}{_\vv{{cap}_e}}$) as given in \eqref{norm}. 
The block height of the channel's funding transaction is taken as the channel age.  

The calculation for the normalized value $\vv{n}_{D}(v)$ of a real number $v$ within the range $D$ is as follows:
\begin{equation}\label{norm}
\begin{aligned}
    \vv{n}_{D}(v) &= 0.00001 \\
    &\quad + 0.99998 \cdot \frac{\min(\max(\min D, v), \max D) - \min D}{\max D - \min D},
\end{aligned}
\end{equation}
where $\text{min} D$ and $\text{max} D$ represent the minimum and maximum values in the range $D$, respectively.
\subsubsection{Case 2: Heuristics constants (without logarithm)}

In this case, the weight of an edge takes into account the risk cost, a virtual failure cost, and channel-wise success probability, alongside the fees and hop cost for the channel $e$:
\begin{equation}\label{case2:eclair}
\begin{aligned}
    \vv{weight(e)} &= \vv{fee}_e + \vv{HopCost}_e + \vv{RiskCost}_e \\
    &\quad + \left(\frac{\vv{FailureCost}_e}{P_e}\right).
\end{aligned}
\end{equation}
\subsubsection{Case 3: Heuristics constants (with logarithm)}

This case is similar to case 2, but here, the weight function incorporates the logarithm of success probability, such that
\begin{equation}\label{case3:eclair}
\begin{aligned}
    \vv{weight(e)} &=  \vv{fee}_e + \vv{HopCost}_e + \vv{RiskCost}_e \\
    &\quad - \vv{FailureCost}_e \cdot \log(P_e).
\end{aligned}
\end{equation}

In cases 2 and 3, success probability is determined by computing the ratio of the amount to the capacity, as given below,
\begin{equation}\label{prob:eclair}
    P_e =  1 - \frac{\vv{amt}_e}{\vv{cap}_e}.
\end{equation}
In routing, as mentioned in \Cref{sec:designelements}, there is a constant risk of funds being locked in HTLC. This risk is represented by the heuristic constant in Eclair, denoted as $\vv{lockedFundsRisk}$, with a preset value of $10^{-8}$. Therefore, $\vv{RiskCost}_e$ is calculated to manage the influence of timelock constraints and is given as:
\begin{equation}\label{riskcost}
    \vv{RiskCost}_e =  \vv{amt}_e \cdot \timelockdelta \cdot \vv{lockedFundsRisk}.
\end{equation}
Failure cost represents a virtual cost associated with failed payment attempt, similar to the use of attempt cost in LND \eqref{attemptcost:lnd}, defined as:
\begin{equation}\label{failurecost}
\begin{aligned}
    \vv{FailureCost}_e &= \vv{BaseFailureCost} \\
    &\quad + \vv{amt}_e \cdot \vv{FailureCostRate},
\end{aligned}
\end{equation}
where $\vv{BaseFailureCost}$ and $\vv{FailureCostRate}$ are given with default values 2000 msat and 500 msat respectively.

Similarly, $\vv{HopCost}_e$, a virtual cost to optimize the path length, is defined as:
\begin{equation}\label{hopcost}
    \vv{HopCost}_e =  \vv{BaseHopCost} + \vv{amt}_e \cdot \vv{HopCostRate}.
\end{equation}
The default values of $\vv{BaseHopCost}$ and $\vv{HopCostRate}$ are both zero. A higher $\vv{HopCost}_e$ strongly penalizes longer paths. Fine-tuning $\vv{HopCost}_e$ can be considered to optimize for path length. The fees associated with per-hop charges and failed attempts are not actually spent. They function as incentives to favor shorter paths with greater probability of success. 

Eclair integrates its weight function into Yen's $K$-shortest path algorithm for pathfinding, with the default number of paths set to $K = 3$. After finding paths, it validates constraints such as the timelock limit $\sum_{e \in p} \timelockdelta \leq \vv{2016}$, the fee limit $\sum_{e \in p} \vv{fee}_e \leq \vv{User\ Defined\ Max\_Fee\_Limit}$, and the path length limit $\sum_{e \in p} 1 \leq \vv{20}$.

\section{Evaluation}\label{sec:evaluation}
In this section, we present the simulation model designed to evaluate the empirical performance of four LN clients (LND v0.17.4-beta, CLN v24.02.1, LDK v0.0.120, and Eclair v0.10.0), including their variations outlined in \Cref{sec:client:weights}. We also discuss our experiment methodology, results, and observations based on various evaluation metrics.

\subsection{Simulation Model}\label{subsec:simulation_model}
Our simulation model is developed using Python 3.11 for all experiments. The source code is available on GitHub at: \url{https://github.com/sindurasaraswathi/LN-PathFinding}. The Lightning Network graph is constructed from the dataset \cite{key}, and all graph-related operations are performed using the Networkx library \cite{networkxNetworkXx2014}. The utilized Lightning Network snapshot \cite{key} comprises 13,129 nodes and 57,773 bi-directional channels, providing channel-specific information such as short channel ID, channel ID, capacity, base fee, proportional fee, minimum and maximum HTLC, and CLTV delta.

All relevant information required for pathfinding and used in weight functions is included as attributes of the channel in the generated graph. Channel age is extracted using the short channel ID. Consistent with the available information of an arbitrary LN node operator who is not engaged in balance probing, the dataset \cite{key} does not contain information about channel balances, which is only known to the two participating nodes of channel. To simulate the channel balances, we use two random models for the balances. The first model is built by uniformly choosing channel balances between zero and the channel capacity. The second graph model is constructed by sampling channel balances from a bimodal distribution with density $\vv{P}(x)$ proportional to 
\begin{equation} 
\label{eq:bimodal:simulation}
    \vv{P}(x) \sim e^{\frac{\vv{-x}}{\vv{s}}} + e^{\frac{\vv{x-cap}_e}{\vv{s}}}
\end{equation}
on $[0, \vv{cap}_e]$, where $\vv{s}$ is selected as a fixed fraction of the channel capacity, specifically $\vv{s} = \frac{\vv{cap}_e}{10}$. We refer to the first and second graphs as the uniform distribution graph and the bimodal distribution graph, respectively, throughout this section.

In general, the Lightning Network comprises of nodes with varying levels of connectivity, determined by factors such as capacity and the number of channels. For instance, as mentioned in \cite{antonopoulos2021mastering}, there are nodes operated by prominent merchants, which are typically well-funded and highly connected with numerous channels. Conversely, there are nodes with moderate connectivity, possessing adequate capacity and channels. Additionally, the network includes nodes that are poorly connected, lacking in channels and funds.

Understanding the routing behavior of nodes across these different connectivity levels is also crucial. Hence, we distinguish \emph{well-connected}, \emph{fairly connected}, and \emph{poorly connected} nodes based on the total capacity across all channels originating from a node and the number of channels a node is participating in. Specifically, nodes with a total capacity greater than or equal to $10^6$ sats and more than 5 channels are called \emph{well-connected}, those with a total capacity falling between $10^4$ and $10^6$ sats and more than 5 channels are called \emph{fairly connected}, and nodes with $5$ or fewer channels are designated as \emph{poorly connected}, regardless of their total channel capacity. The statistics on node connectivity, including the number of nodes and channels, as well as the channel capacity for each node category, are presented in \Cref{table:node:stat}.

\begin{table*}
  \caption{Node connectivity statistics}
  \label{table:node:stat}
  \centering
  {
  \begin{tabular}{llll}
    \toprule
Connectivity Level & Node count & Median channel count & Median capacity \\
\midrule
Well-connected & 3476 & 12 & 24616394 \\
Fairly-connected & 53 & 6 & 535000 \\
Poorly-connected & 9600 & 1 & 125000 \\
\bottomrule
  \end{tabular}
  }
\end{table*}
We conduct several experiments to compare the empirical performance of the four Lightning clients and their variations. Initially, using the uniform distribution graph, simulations are performed for 10,000 transactions, with source and receiver nodes randomly selected from all nodes in the graph. Transaction amounts are uniformly chosen between $1$ and the minimum of the maximum values of all outgoing channel balances of the source node and all local incoming balances of the receiver node. By using this upper bound, we can select payment amounts that are practically feasible for the source node to send and for the receiver node to receive.

We use the $\_dijkstra$ method from the NetworkX library to execute the Dijkstra's algorithm for CLN and LDK, and adapted $\_dijkstra$ method for LND as discussed in \Cref{sec:beyond:shortest:path}, supplying the method with an appropriate weight function. For finding Yen's K shortest paths in the case of Eclair, we adapt the $shortest\_simple\_paths$ method from the NetworkX library. Upon determining a path for a given Lightning client and transaction, our model simulates the routing of the payment amount through the determined path. The routing is regarded a failure if no acceptable path could be found or if any of the intermediary nodes has insufficient balances.

In our experiments, we study and compare a total of eight variants of LN clients: LND-ap, LND-bm, CLN, LDK-un, LDK-bm, Eclair1, Eclair2, and Eclair3. LND-ap and LND-bm represent the LND implementation with $Apriori$ \eqref{prob:apriori:lnd} and $Bimodal$ \eqref{prob2:bimodal:lnd} estimators, respectively, for channel-wise success probability calculation. The LDK implementation with $Uniform$ \eqref{prob:ldk} and $Bimodal$ \eqref{prob:ldk_bimodal} estimators for channel-wise success probability is represented as LDK-un and LDK-bm, respectively. Similarly, Eclair1, Eclair2, and Eclair3 denote three distinct cases of Eclair: heuristics ratios, heuristics constants (without logarithm) and heuristics constants (with logarithm), respectively, as discussed in \Cref{sec:client:weights}.
Building on the existing LND implementation, we propose and study a new weight function by assuming a uniform liquidity distribution for calculating channel-wise success probability, defined as 
\begin{equation}\label{prob:lnd-un}
P_e = \frac{\vv{cap_e - amt_e}}{\vv{cap_e}}.
\end{equation}
The expression \eqref{prob:lnd-un} is used to compute the path success probability, which is subsequently incorporated into the LND weight function \eqref{weight:lnd}. We refer to this implementation as LND-un. In the experiments, we use the default values for all constants in the weight function of each LN client variant, as mentioned in \Cref{sec:client:weights}.

\subsection{Simulation Results}
\subsubsection{LN Graph with Uniform Balance Distribution}
We investigate the behavior of each variant of the LN clients across various transaction amount bins, spanning the range from 1 to $10^8$. As a first experiment, whose results are reported in \Cref{table:successpercent,table:metrics,table:feeratio_uniform}, we sample channel balances from the uniform distribution. 

We observe in \Cref{table:successpercent} that for the uniform distribution graph, LND-un exhibits the highest success rate across most of the amount bins compared to all other existing LN client variants. Among the eight existing LN client variants, Eclair3 demonstrates high success rates. LND-ap and LND-bm show similar performance levels, with LND-bm slightly outperforming LND-ap in success rates. CLN’s success rate remains competitive for larger amounts and outperforms LND-ap, while LDK-un and LDK-bm perform moderately for small and lower mid-range amounts; however, their success rates experience a significant decline for larger amounts. Eclair1 and Eclair2 demonstrate similar performance levels, with higher success rates observed for larger payment amounts compared to CLN, LDK variants, and LND variants.

While the success rate, or payment reliability, remains a crucial metric in pathfinding algorithm design, it is not the sole consideration. Factors such as the fee incurred, the path length, and the total timelock or latency along the path also play significant roles. Consequently, we provide also alternative performance metrics of LN client variants in
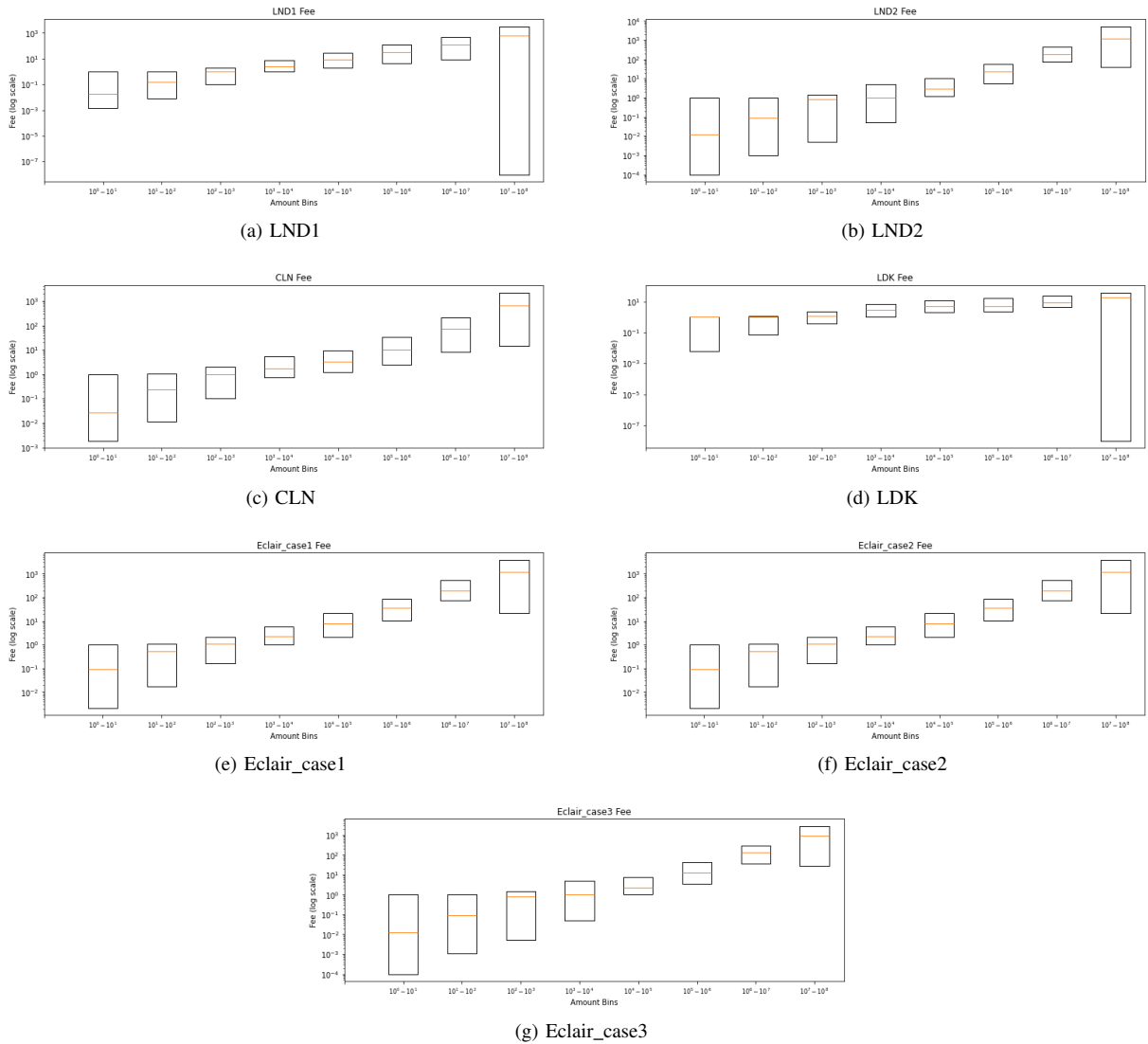
\Cref{table:metrics}, which includes a fee ratio value, average path length, and average timelock, for which information across all amount bins of \Cref{table:successpercent} is aggregated. To ensure a fair comparison, we exclusively considered transactions for which routing succeeded in five or more routing client variants. The fee ratio of \Cref{table:metrics} is calculated as the median of the fraction of incurred fees and the transaction amount, with the median taken across all transactions under consideration. To understand the distribution of fees across transaction amount bins, we also report the median fee ratio for each bin in \Cref{table:feeratio_uniform}.

Among the considered LN clients, Eclair3 demonstrates effective overall performance in terms of fee along the path, providing the lowest fee ratio compared to other clients. LND-un, in addition to achieving a high success rate, yields paths with lower fees but relatively longer path lengths. LND-bm provides paths with lower fees, moderate path lengths, and significantly higher latency. LND-ap, on the other hand, offers paths with significantly higher fees but shorter path lengths and lower latency. Among the routing clients, LDK-un and LDK-bm provide paths with high fees, with LDK-bm exhibiting the highest fee ratio. However, for larger payment amounts, as seen in \Cref{table:feeratio_uniform}, LDK-un and LDK-bm offer the lowest fee ratios compared to all other clients. CLN emerges as the top performer in terms of total timelock, offering paths with the lowest latency for routing payment amounts. Eclair1 and Eclair2 perform the best in terms of path length, while Eclair3 exhibits the highest path length and latency.

\begin{table*}
  \caption{Success rates (in $\%$) of Lightning Network clients across payment amount ranges, uniform balance distribution}
  \label{table:successpercent}
  \centering
  {
  \begin{tabular}{llllllllll}
    \toprule
Amount Bins & LND-ap & LND-bm & LND-un & CLN & LDK-un & LDK-bm & Eclair1 & Eclair2 & Eclair3 \\
\midrule
$10^0-10^1$ & \textbf{98.103} & \textbf{98.103} & \textbf{98.103} & \textbf{98.103} & 98.024 & 98.024 & \textbf{98.103} & \textbf{98.103} & \textbf{98.103} \\
$10^1-10^2$ & \textbf{98.139} & \textbf{98.139} & \textbf{98.139} & \textbf{98.139} & \textbf{98.139} & \textbf{98.139} & \textbf{98.139} & \textbf{98.139} & \textbf{98.139} \\
$10^2-10^3$ & 97.598 & \textbf{97.838} & 97.678 & 97.518 & \textbf{97.838} & 97.758 & 97.598 & 97.598 & \textbf{97.838} \\
$10^3-10^4$ & 88.480 & 90.080 & \textbf{90.320} & 89.360 & 89.040 & 89.200 & 89.280 & 89.280 & 90.400 \\
$10^4-10^5$ & 69.280 & 72.800 & \textbf{74.800} & 71.040 & 68.480 & 69.040 & 72.400 & 72.400 & 74.240 \\
$10^5-10^6$ & 50.160 & 59.040 & \textbf{66.720} & 54.480 & 46.560 & 45.440 & 61.360 & 61.360 & 65.280 \\
$10^6-10^7$ & 33.200 & 37.040 & \textbf{47.680} & 36.720 & 22.960 & 21.680 & 44.240 & 44.240 & 45.040 \\
$10^7-10^8$ & 27.840 & 27.840 & \textbf{36.560} & 30.400 & 16.960 & 16.240 & 34.800 & 34.800 & 35.200 \\
\bottomrule
  \end{tabular}
  }
\end{table*}
\begin{table*}
  \caption{Comparison of performance metrics across LN clients, uniform balance  distribution}
  \label{table:metrics}
  \centering
  {
  \begin{tabular}{llllllllll}
    \toprule
Attribute & LND-ap & LND-bm & LND-un & CLN & LDK-un & LDK-bm & Eclair1 & Eclair2 & Eclair3 \\
\midrule
Fee Ratio (median) & 0.0500\% & 0.0200\% & 0.0142\% & 0.0271\% & 0.0550\% & 0.0564\% & 0.0503\% & 0.0503\% & \textbf{0.0116\%} \\
Path Length (avg.) & 4.4505 & 5.4956 & 6.1855 & 4.6342 & 4.5783 & 4.5703 & \textbf{4.4242} & \textbf{4.4242} & 6.4419 \\
Timelock (avg.) & 196.2460 & 232.0691 & 252.3624 & \textbf{186.7123} & 200.0157 & 199.7607 & 195.5912 & 195.5912 & 263.9637 \\
\bottomrule
  \end{tabular}
  }
\end{table*}
\begin{table*}
  \caption{Median fee ratios (in $\%$) incurred by LN clients across transaction amount ranges, uniform balance distribution}
  \label{table:feeratio_uniform}
  \centering
  {
  \begin{tabular}{llllllllll}
    \toprule
Amount Bins & LND-ap & LND-bm & LND-un & CLN & LDK-un & LDK-bm & Eclair1 & Eclair2 & Eclair3 \\
\midrule
$10^0-10^1$ & 0.3148 & \textbf{0.3005} & \textbf{0.3005} & 0.3498 & 10.3963 & 10.3802 & 0.6015 & 0.6015 & 0.3007 \\
$10^1-10^2$ & 0.3050 & \textbf{0.3001} & \textbf{0.3001} & 0.3318 & 1.2588 & 1.2722 & 0.6011 & 0.6011 & \textbf{0.3001} \\
$10^2-10^3$ & 0.2080 & \textbf{0.1277} & \textbf{0.1277} & 0.1957 & 0.3017 & 0.3026 & 0.2628 & 0.2628 & \textbf{0.1277} \\
$10^3-10^4$ & 0.0504 & 0.0180 & \textbf{0.0176} & 0.0302 & 0.0531 & 0.0551 & 0.0422 & 0.0422 & \textbf{0.0176} \\
$10^4-10^5$ & 0.0200 & 0.0091 & 0.0069 & 0.0083 & 0.0100 & 0.0107 & 0.0185 & 0.0185 & \textbf{0.0052} \\
$10^5-10^6$ & 0.0103 & 0.0109 & 0.0064 & 0.0035 & \textbf{0.0016} & \textbf{0.0016} & 0.0105 & 0.0105 & 0.0045 \\
$10^6-10^7$ & 0.0063 & 0.0097 & 0.0059 & 0.0055 & 0.0008 & \textbf{0.0007} & 0.0082 & 0.0082 & 0.0057 \\
$10^7-10^8$ & 0.0055 & 0.0055 & 0.0056 & 0.0055 & \textbf{0.0041} & \textbf{0.0041} & 0.0056 & 0.0056 & 0.0055 \\
\bottomrule
  \end{tabular}
  }
\end{table*}
\begin{table*}
  \caption{Success rates (in $\%$) of Lightning Network clients across payment amount ranges, bimodal balance distribution}
  \label{table:successbimodal}
  \centering
  {
  \begin{tabular}{lllllllllll}
    \toprule
Amount Bins & LND-ap & $\underset{(s=3e5)}{\text{LND-bm}}$ & $\underset{(s=\frac{cap}{10})}{\text{LND-bm}}$ & LND-un & CLN & LDK-un & LDK-bm & Eclair1 & Eclair2 & Eclair3 \\
\midrule
$10^0-10^1$ & \textbf{97.630} & 97.472 & 97.551 & 97.472 & \textbf{97.630} & 97.551 & 97.551 & 97.551 & 97.551 & 97.472 \\
$10^1-10^2$ & \textbf{98.543} & 98.462 & \textbf{98.543} & 98.462 & 98.462 & 98.219 & 98.219 & 98.462 & 98.462 & \textbf{98.543} \\
$10^2-10^3$ & 93.435 & 93.755 & \textbf{94.476} & 94.155 & 93.595 & 92.714 & 92.874 & 93.755 & 93.755 & 94.396 \\
$10^3-10^4$ & 78.080 & 79.360 & \textbf{81.600} & 80.240 & 78.160 & 76.720 & 77.760 & 79.520 & 79.520 & 79.520 \\
$10^4-10^5$ & 51.120 & 55.280 & \textbf{61.600} & 58.400 & 51.520 & 47.040 & 49.360 & 54.640 & 54.640 & 57.280 \\
$10^5-10^6$ & 34.720 & 36.320 & \textbf{45.600} & 40.560 & 33.520 & 21.840 & 23.120 & 38.560 & 38.560 & 37.680 \\
$10^6-10^7$ & 26.960 & 28.160 & \textbf{35.920} & 33.120 & 26.080 & 13.360 & 13.760 & 30.480 & 30.480 & 29.760 \\
$10^7-10^8$ & 21.360 & 21.200 & \textbf{22.480} & 20.800 & 18.480 & 11.520 & 11.200 & 21.840 & 21.840 & 19.520 \\
\bottomrule
  \end{tabular}
  }
\end{table*}

\begin{table*}
  \caption{Comparison of performance metrics across LN clients, bimodal balance distribution}
  \label{table:metricsbimodal}
  \centering
  {
  \begin{tabular}{lllllllllll}
    \toprule
Attribute & LND-ap & $\underset{(s=3e5)}{\text{LND-bm}}$ & $\underset{(s=\frac{cap}{10})}{\text{LND-bm}}$ & LND-un & CLN & LDK-un & LDK-bm & Eclair1 & Eclair2 & Eclair3 \\

\midrule
Fee Ratio (median) & 0.0509\% & 0.0201\% & 0.0202\% & 0.0175\% & 0.0355\% & 0.0817\% & 0.0867\% & 0.0571\% & 0.0571\% & \textbf{0.0148\%} \\
Path Length (avg.) & 4.4528 & 5.6160 & 6.2359 & 6.2138 & 4.6012 & 4.5122 & 4.5026 & \textbf{4.3929} & \textbf{4.3929} & 6.4737 \\
Timelock (avg.) & 196.8925 & 238.4878 & 259.2788 & 257.4595 & \textbf{187.9205} & 199.8414 & 199.9199 & 196.5492 & 196.5492 & 268.5613 \\
\bottomrule
  \end{tabular}
  }
\end{table*}
\begin{table*}
  \caption{Median fee ratios (in $\%$) incurred by LN clients across payment amount ranges, bimodal balance distribution}
  \label{table:feeratio_bimodal}
  \centering
  {
  \begin{tabular}{lllllllllll}
    \toprule
Amount Bins & LND-ap & $\underset{(s=3e5)}{\text{LND-bm}}$ & $\underset{(s=\frac{cap}{10})}{\text{LND-bm}}$ & LND-un & CLN & LDK-un & LDK-bm & Eclair1 & Eclair2 & Eclair3 \\
\midrule
$10^0-10^1$ & 0.3068 & \textbf{0.3000} & \textbf{0.3000} & \textbf{0.3000} & 0.3211 & 10.3301 & 10.3301 & 0.6729 & 0.6729 & \textbf{0.3000} \\
$10^1-10^2$ & 0.3050 & \textbf{0.3001} & \textbf{0.3001} & \textbf{0.3001} & 0.3560 & 1.3145 & 1.3265 & 0.7098 & 0.7098 & \textbf{0.3001} \\
$10^2-10^3$ & 0.2283 & \textbf{0.1190} & \textbf{0.1190} & \textbf{0.1190} & 0.2283 & 0.3007 & 0.3005 & 0.2756 & 0.2756 & \textbf{0.1190} \\
$10^3-10^4$ & 0.0457 & 0.0188 & 0.0200 & 0.0184 & 0.0310 & 0.0486 & 0.0500 & 0.0405 & 0.0405 & \textbf{0.0182} \\
$10^4-10^5$ & 0.0200 & 0.0080 & 0.0101 & 0.0054 & 0.0079 & 0.0098 & 0.0100 & 0.0192 & 0.0192 & \textbf{0.0044} \\
$10^5-10^6$ & 0.0128 & 0.0124 & 0.0094 & 0.0062 & 0.0031 & \textbf{0.0013} & \textbf{0.0013} & 0.0112 & 0.0112 & 0.0033 \\
$10^6-10^7$ & 0.0078 & 0.0100 & 0.0106 & 0.0063 & 0.0039 & 0.0005 & \textbf{0.0004} & 0.0108 & 0.0108 & 0.0057 \\
$10^7-10^8$ & 0.0056 & 0.0056 & 0.0100 & 0.0100 & 0.0055 & \textbf{0.0025} & \textbf{0.0025} & 0.0056 & 0.0056 & 0.0056 \\
\bottomrule
  \end{tabular}
  }
\end{table*}
\begin{table*}
  \caption{Success rates (in $\%$) of LND-bm across various liquidity broadening scales ($s$) for the bimodal distribution graph}
  \label{table:LND-bm_scales}
  \centering
  {
  \begin{tabular}{llllllllllllllll}
    \toprule
\diagbox[width=5em,height=2em]{Bins}{s} & 3e5 &$\frac{cap}{2}$ & $\frac{cap}{4}$ & $\frac{cap}{6}$ & $\frac{cap}{8}$ & $\frac{cap}{10}$ & $\frac{cap}{12}$ & $\frac{cap}{14}$ & $\frac{cap}{50}$ & $\frac{cap}{100}$ & $\frac{cap}{10^3}$ & $\frac{cap}{10^4}$ & $\frac{cap}{10^6}$ & $\frac{cap}{10^8}$\\
\midrule

$10^4-10^5$ &55.44 &58.32 &59.44 &59.92 &60.80 &60.88 &61.36 &61.60 &\textbf{63.6} &63.12 &57.28 &53.6 &52.16 &52.16 \\ 
$10^5-10^6$ &38.88	&46.80 &47.20 &47.68 &48.56 &50.08 &\textbf{50.40} &50 &47.52 &44.40	&37.20 &36.16 &36.16 &36.16 \\ 
$10^6-10^7$ &28.16 &31.28 &32.56 &34 &34.16	&34.56 &35.36 &\textbf{35.44} &31.52	&29.60 &26.96 &26.56 &26.48 &26.48 \\ 
\bottomrule
  \end{tabular}
  }
\end{table*}

\begin{table*}
  \caption{Success rates (in $\%$) for amounts in range 10-100 sats across different node category combinations for source and receiver}
  \label{table:nodecategories1}
  \centering
  {
  \begin{tabular}{llllllllll}
    \toprule
    Source-Receiver & LND-ap & LND-bm & LND-un & CLN & LDK-un & LDK-bm & Eclair1 & Eclair2 & Eclair3 \\ 
\midrule
Well-Well   & \textbf{100}    & \textbf{100}    & \textbf{100}    & \textbf{100}    & \textbf{100}    & \textbf{100}    & \textbf{100}    & \textbf{100}    & \textbf{100}    \\
Well-Fair   & \textbf{100}    & \textbf{100}    & \textbf{100}    & \textbf{100}    & 99.394 & 99.394 & 99.394 & 99.394 & \textbf{100}    \\
Well-Poor   & \textbf{100}    & \textbf{100}    & \textbf{100}    & \textbf{100}    & \textbf{100}    & \textbf{100}    & \textbf{100}    & \textbf{100}    & \textbf{100}    \\
Fair-Well   & \textbf{100}    & \textbf{100}    & \textbf{100}    & \textbf{100}    & \textbf{100}    & \textbf{100}    & \textbf{100}    & \textbf{100}    & \textbf{100}    \\
Fair-Fair   & \textbf{100}    & \textbf{100}    & \textbf{100}    & \textbf{100}    & \textbf{100}    & \textbf{100}    & \textbf{100}    & \textbf{100}    & \textbf{100}    \\
Fair-Poor   & 98.193 & \textbf{98.795} & \textbf{98.795} & \textbf{98.795} & \textbf{98.795} & \textbf{98.795} & \textbf{98.795} & \textbf{98.795} & \textbf{98.795} \\
Poor-Well   & \textbf{100}    & \textbf{100}    & \textbf{100}    & \textbf{100}    & \textbf{100}    & \textbf{100}    & \textbf{100}    & \textbf{100}    & \textbf{100}    \\
Poor-Fair   & 95.783 & 96.988 & 96.988 & 96.386 & 96.988 & 96.988 & 96.988 & 96.988 & \textbf{97.590} \\
Poor-Poor   & \textbf{99.187} & \textbf{99.187} & \textbf{99.187} & \textbf{99.187} & \textbf{99.187} & \textbf{99.187} & \textbf{99.187} & \textbf{99.187} & \textbf{99.187} \\
\bottomrule
  \end{tabular}
  }
\end{table*}
\begin{table*}
  \caption{Success rates (in $\%$) for amounts in the range of $10^3$-$10^4$ sats across different node category combinations for source and receiver}
  \label{table:nodecategories2}
  \centering
  {
  \begin{tabular}{llllllllll}
    \toprule
    Source-Receiver & LND-ap & LND-bm & LND-un & CLN & LDK-un & LDK-bm & Eclair1 & Eclair2 & Eclair3 \\ 
\midrule
Well-Well   & 99.2    & 99.2    & 99.2    & \textbf{100}    & 99.2    & 99.2    & \textbf{100}    & \textbf{100}    & 99.2    \\
Well-Fair   & 89.820  & 94.012  & \textbf{94.611 } & 92.216 & 91.018 & 90.419 & 93.413 & 93.413 & 94.012  \\
Well-Poor   & 90.4    & \textbf{91.2}    & 89.6    & 90.4   & 88.8   & 89.6   & \textbf{91.2}   & \textbf{91.2}   & 88.8    \\
Fair-Well   & \textbf{99.401}  & 97.605  & 95.210  & 97.605 & 89.820 & 91.617 & \textbf{99.401} & \textbf{99.401} & 95.210  \\
Fair-Fair   & 85.629  & 92.814  & \textbf{94.012}  & 87.425 & 86.826 & 88.623 & 86.228 & 86.228 & 92.814  \\
Fair-Poor   & 88.623  & 92.814  & \textbf{93.413}  & 89.820 & 83.234 & 83.832 & 88.024 & 88.024 & 92.814  \\
Poor-Well   & 93.6    & 93.6    & 93.6    & 93.6   & 92.8   & 92     & 93.6   & 93.6   & \textbf{94.4}    \\
Poor-Fair   & 81.437  & 88.024  & 88.623  & 83.832 & 86.826 & \textbf{89.820} & 85.629 & 85.629 & 88.623  \\
Poor-Poor   & 83.2    & 84.8    & \textbf{86.4}    & 85.6   & 84     & 84     & 84.8   & 84.8   & \textbf{86.4}    \\
\bottomrule
  \end{tabular}
  }
\end{table*}
\begin{table*}
  \caption{Success rates (in $\%$) for amounts in the range of $10^5$-$10^6$ sats across different node category combinations for source and receiver}
  \label{table:nodecategories3}
  \centering
  {
  \begin{tabular}{llllllllll}
    \toprule
    Source-Receiver & LND-ap & LND-bm & LND-un & CLN & LDK-un & LDK-bm & Eclair1 & Eclair2 & Eclair3 \\ 
\midrule
Well-Well   & 64     & 73.6   & \textbf{77.6}   & 67.2   & 52.8   & 50.4   & 74.4   & 74.4   & 75.2   \\
Well-Fair   & 27.711 & \textbf{31.928} & 31.325 & 27.108 & 24.699 & 25.904 & \textbf{31.928} & \textbf{31.928} & 31.325 \\
Well-Poor   & 56     & 57.6   & \textbf{63.2}   & 56.8   & 51.2   & 48     & 59.2   & 59.2   & 60     \\
Fair-Well   & 31.325 & 33.133 & 32.530 & 33.133 & 24.699 & 23.494 & \textbf{33.735} & \textbf{33.735} & \textbf{33.735} \\
Fair-Fair   & 14.458 & 24.096 & 23.494 & 17.470 & 18.072 & 15.060 & 23.494 & 23.494 & \textbf{25.301} \\
Fair-Poor   & 33.735 & 37.349 & \textbf{37.952} & 33.133 & 21.084 & 21.084 & 36.145 & 36.145 & 34.337 \\
Poor-Well   & 46.4   & 54.4   & 59.2   & 52.8   & 36.8   & 36.8   & 57.6   & 57.6   & \textbf{60}     \\
Poor-Fair   & 30.120 & 34.337 & \textbf{37.349} & 27.108 & 24.096 & 25.301 & 34.337 & 34.337 & 31.325 \\
Poor-Poor   & 36     & 43.2   & 46.4   & 40     & 30.4   & 31.2   & \textbf{47.2}   & \textbf{47.2}   & 46.4   \\
\bottomrule
  \end{tabular}
  }
\end{table*}

\subsubsection{LN Graph with Bimodal Balance Distribution}
To obtain a more complete picture of the LN clients' performance given channel balances that are possibly more realistic in the real LN graph, we report the results of a similar simulation on the bimodal distribution graph in \Cref{table:successbimodal,table:metricsbimodal,table:feeratio_bimodal}, in which the channel liquidities are sampled from \eqref{eq:bimodal:simulation} with $s=\frac{\vv{cap}_e}{10}$, so that the liquidities are predominantly concentrated on either side of the channel. 

In addition to LND-bm, which uses the default value of ${3\cdot 10^{5}\;\vv{sats}}$ for the liquidity broadening scale $\vv{s}$ in \eqref{prob2:bimodal:lnd}, we also include simulations for LND-bm with $\vv{s}$ set to $\frac{\vv{cap_e}}{10}$ in \eqref{prob2:bimodal:lnd}, the same scale we used for sampling channel balances in the bimodal distribution graph.
 
We observe in \Cref{table:successbimodal}, that for an LN graph with bimodal balance distribution, LND-bm with broadening scale parameter $\vv{s}=\frac{\vv{cap_e}}{10}$ yields the highest success rates across all amount bins except for micropayments between $1$ and $10$ sats. In contrast, LND-bm with the default value of $\vv{s}$ shows comparatively lower success rates. Similar to the uniform distribution case, LND-ap and the default LND-bm exhibit similar performance levels. LND-un performs competitively, achieving the second-highest success rates among LN clients. CLN performs moderately, while LDK-un and LDK-bm deliver the lowest success rates across the amount bins. Eclair1 and Eclair2 demonstrate competitive success rates, slightly outperforming Eclair3, in contrast to their performance in the uniform distribution case.

In comparison to the uniform distribution case results of \Cref{table:successpercent}, success rates in the bimodal distribution graph are significantly lower. We attribute this decline in success rates to the bimodal distribution of channel liquidity, which inherently limits the routing ability of a given channel in one of the two directions unless the payment amount is very small, unlike in the uniform distribution case. As the payment amount increases, it becomes challenging to identify reliable routes for payment delivery due to the scarcity of available liquidity. 

However, \Cref{table:metricsbimodal} shows only a slight increase in the fee ratio across all LN clients under the bimodal distribution scenario. Even in the presence of the bimodal distribution graph, Eclair3 continues to achieve the lowest fee ratio. In addition, Eclair1 and Eclair2 display the shortest path lengths, while CLN maintains the lowest overall timelock.  \Cref{table:feeratio_bimodal} implies that LDK-un and LDK-bm, despite their high overall fee ratios, provide paths with the lowest fees for large payment amounts, similar to the uniform distribution case.

\subsubsection{Choice of Broadening Scale $s$ in LND-bm}
Since enhanced success rates are observed for the bimodal distribution graph when using $\frac{\vv{cap}_e}{10}$ for $\vv{s}$ in \eqref{prob2:bimodal:lnd}---the same scale used during channel liquidity sampling---compared to the default value, we conduct an ablation study to further investigate this effect, which we report in \Cref{table:LND-bm_scales}. Specifically, 3750 transactions are simulated across bins ranging from $10^4$ to $10^7$, with $\vv{s}$ values varied as shown in \Cref{table:LND-bm_scales}. The simulation results reveal that the success rates improve when $\vv{s}$ aligns closely with the scale used for channel liquidity sampling. These findings suggest that tuning the liquidity broadening scale $\vv{s}$ in \eqref{prob2:bimodal:lnd} based on historical channel liquidity data could significantly enhance payment reliability.

\subsubsection{Effect of Node Connectivity Levels}
To study the effect of node connectivity levels in routing payments, we simulate transactions on the uniform distribution graph, by selecting nodes from well, fairly, and poorly connected node categories, resulting in a total of 9 experiments, each consisting of 1,000 transactions. These experiments cover all possible combinations of node categories by selecting source and destination nodes from the three connectivity levels.

In all possible node category combinations for selecting source and receiver nodes, routing clients achieve varying success rates in different bins, as illustrated in \Cref{table:nodecategories1}, \Cref{table:nodecategories2} and \Cref{table:nodecategories3}. 

Since we classify nodes based on both the number of channels and their total capacity, fairly connected nodes, with capacities ranging between $10^4$ and $10^6$, can only accommodate transaction amounts less than $10^6$, resulting in bins ranging from $1$ to $10^6$. In contrast, well and poorly connected nodes may have capacities exceeding $10^6$, enabling transaction amounts from bins ranging from $1$ to $10^8$. Therefore, to ensure a meaningful comparison of results, we compare success rates within each bin ranging from $1$ to $10^6$ separately. From \Cref{table:nodecategories1}, we observe that for small payment amounts (10-100 sats), the success rates are high across all node category combinations. However, notably, for combinations involving poorly connected nodes, the success rates are relatively lower.

Overall from \Cref{table:nodecategories2} and \Cref{table:nodecategories3}, across all routing clients, we observe that well-connected nodes exhibit high success rates, along with low fees, low delay, and short path lengths. Fairly connected nodes, on the other hand, provide lower success rates, charge moderate fees, with moderate delay and path length. In case of poorly connected nodes, the success rate surpasses that of fairly connected nodes. This is attributed to poorly connected nodes having substantial capacity, enabling successful payment routing despite having fewer channels. However, it should be noted that poorly connected nodes typically impose high fees, experience high delay, and have longer path lengths which are undesirable for efficient routing.

From the results of these connectivity level routing evaluation experiments, we find that LND-un delivers the highest success rates across multiple node categories, while LND-ap performs poorly when the receiving node is fairly connected. LDK-un and LDK-bm demonstrate low reliability across most node combinations. CLN exhibits moderate reliability; however, in situations where nodes are poorly connected, CLN performs competitively. Eclair1, Eclair2, and Eclair3 emerge as strong contenders due to their high reliability rates across various node categories.

\subsection{Analysis of Results}
Analyzing the results from all our experiments and interpreting them using the weight functions from \Cref{sec:client:weights}, several insights can be inferred. LND-un, formulated through a modification of LND's channel success probability estimate, results in payment paths with high success rates. While the incorporation of more complex models that also leverage historical or external information about channel balances is beyond the scope of our study, we think that this indicates that reliability can be further improved by an adept modeling of success probabilities in a LN client's weight function.

Despite the relatively realistic modeling of success probabilities in LND-bm, we observed that LND-bm exhibits relatively moderate reliability with default constants in the weight function. Our ablation study of \Cref{table:feeratio_bimodal} suggests that success rates of LND-bm can potentially be improved by fine-tuning the liquidity broadening scale $\vv{s}$ in \eqref{prob2:bimodal:lnd}.

LND-ap's lower success rate can be attributed to the $Apriori$ estimation of the success probability, which assumes a base success likelihood of 0.6. This assumption, along with default values for the constants $\vv{c_o}$ and $\vv{s_o}$ in \eqref{prob:apriori:lnd}, might not yield reliable paths. Moreover, this model takes into account historical payment attempts to adjust the channel-wise success probability as described in \Cref{weight:lnd}. However, in our simulation, we do not consider historical attempts when computing the channel success probabilities, which might have contributed to the low success rate.

By analyzing the different variants of Eclair's pathfinding module, we see that the weight function \eqref{case3:eclair} used by Eclair3 reveals a tendency to favor paths that balance high reliability and low fees. This trend is consistent with the observed outcomes, indicating that Eclair3 consistently attempts to provide paths characterized by both higher reliability and lower fees.
Finally, the decline in reliability for larger amounts in both LDK-un and LDK-bm can be attributed to the weight function \eqref{weight:ldk},  which prioritizes paths with lower fees over higher reliability. The higher fee ratio for small payment amounts in LDK may result from the weight function favoring paths with lower HTLC minimums rather than lower fees as illustrated in \Cref{subsec:LDK_weight}. Moreover, LDK employs historical liquidity tracking to estimate upper and lower bounds of channel liquidity. Since our simulations do not incorporate historical data, this omission may have contributed to the low success rates observed in LDK. Incorporating historical data has a clear potential to further  improve its success rates.

\section{Limitations}\label{limitations}
In our simulation model, we do not consider the historical payment attempts to update the channel-wise success probability. In the event of failure, the success probability of the failed channel is not taken to zero, as it is in LND-ap.

Furthermore, our model neither has the capability to keep track of historical liquidity estimations like channel upper bound and lower bounds, as in LDK-un and LDK-bm, nor does it have the capability to record the previous success and failure amounts in the channel, as is the case in LND-bm. 

Our model does not consider most side constraints (see \Cref{table:side_constraints}) during pathfinding, as these change the complexity class of the pathfinding problem, cf. \Cref{sec:beyond:shortest:path}. Specifically, we do not consider constraints on timelock, path length, and fees.

This limitation may introduce slight variations in the results observed in real Lightning Network routing. Additionally, we used default values provided by the LN clients' open-source implementations for design elements that are constants. Varying these values could yield different results.

\section{Conclusion}\label{conclusion}
In this study, we provide transparency into the pathfinding modules of the four prominent Lightning Network clients LND, CLN, LDK, and Eclair. Using simulated channel liquidities and payment attempts, we further assess their empirical performance across various metrics including success rate, fee ratio, path length, and timelock. Our experiments reveal that Eclair demonstrates superior success rates and distinguishes itself by offering paths with low fees. This performance can be attributed to its weight function, which is designed to trade off between high reliability and low fees. Although LND provides moderate reliability with its default settings, it achieves higher success rates when these settings are fine-tuned. LDK exhibits suboptimal fee performance for smaller payments; however, for larger payment amounts, it provides paths with the lowest fees. Additionally, CLN stands out for offering paths with minimal timelock, consistent with its weight function, which is designed to prioritize timelock minimization.

We find that the connectivity level of the nodes significantly influences the selection of an appropriate path. Depending on the node's connectivity level and the desired performance metric optimization, the selection of the appropriate weight function or routing client becomes crucial.

Furthermore, we observe that certain path cost modeling, while reasonable from a modeling perspective, can harm the performance of the pathfinding algorithm, and simple modifications to standard shortest path algorithms such as Dijkstra's algorithm do not always return the expected, optimal solution, in particular, in the case of LND's cost function. We also observe that the imposition of side constraints (see \eqref{eq:sideconstraint} and \Cref{table:side_constraints}) imposes potentially severe difficulties for the pathfinding process, as the underlying problem becomes a constrained shortest-path problem, which is $\mathcal{NP}$-complete and thus cannot be guaranteed to be solved optimally in practice for all but small graph instances.

We hope that the insights of this study motivate future developments in designing better weight functions which, dependent on the desired properties of the resulting path, can deliver better trade-offs between payment reliability, routing fees and other desired properties. We expect that there is room for substantial improvements in future work. From an algorithmic side, we conclude that it is worthwhile to consider more sophisticated algorithms than Dijkstra's algorithm (or its ad-hoc modifications) to improve pathfinding both in terms of computational efficiency (which corresponds to payment latency) and solution quality (which corresponds to routing fees and payment reliability). 

Finally, while beyond the scope of this study, we note that the insights gained in this study are also relevant for future improvements in multi-part payment pathfinding algorithms, as those algorithms are based on the solution of minimum cost flow problems, which are a generalization of shortest path problems. The choice of an appropriate channel-specific weight/cost function likewise significantly influences the solution quality in multi-part pathfinding. 

\section*{Acknowledgements}
The authors thank Ren\'e Pickhardt, Matt Corallo (BlueMatt) and bitromortac for their helpful comments and feedback on our manuscript. The authors are supported by the National Science Foundation under the grant NSF-CIF-2348430.

\bibliographystyle{unsrt}
\bibliography{reference}
\end{document}